\documentclass[aps,twocolumn,prb,floatfix]{revtex4-2}
\usepackage{epsfig}
\usepackage{dcolumn}
\usepackage{bm}
\usepackage{amsmath}
\usepackage{tikz}
\usepackage{graphicx, graphics, color}
\usepackage{natbib}

\usepackage[colorlinks=true,citecolor=blue]{hyperref}

\ifpdf
	\DeclareGraphicsExtensions{.pdf,.png,.jpg,.mps}
\else
	\DeclareGraphicsExtensions{.eps}
\fi

\newcommand{\be}{\begin{equation}}
\newcommand{\ee}{\end{equation}}
\newcommand{\beq}{\begin{eqnarray}}
\newcommand{\eeq}{\end{eqnarray}}

\graphicspath{{./figs/}}

\begin{document}

\title{Charge and heat transport through quantum dots with local and correlated-hopping interactions}

\author{Ulrich Eckern}
\affiliation{Institute of Physics, University of Augsburg, 86135 Augsburg, Germany}

\author{Karol I.\ Wysoki\'{n}ski}
\affiliation{Institute of Physics, M.~Curie-Sk{\l}odowska University, pl.~M.~Curie-Sk{\l}odowskiej 1, 20-031 Lublin, Poland}


\begin{abstract}
The transport properties of junctions composed of a central region tunnel-coupled to external electrodes 
are frequently studied within the single-impurity Anderson model with Hubbard on-site interaction. 
In the present work, we supplement the model with an important ingredient, namely the charge-bond interaction, 
also known as correlated or assisted hopping. 
Correlated hopping enters the second-quantised Hamiltonian, written in the Wannier representation, as 
an off-diagonal many-body term. Using the equation of motion
technique, we study the effect of the correlated hopping on the spectral and transport 
characteristics of a two-terminal quantum dot. Two different Green functions (GFs) appear: one of them
describes the spectral properties of the quantum dot, the other the transport properties of the system.  
The calculation of the transport GF requires the knowledge of the spectral one. We use
decoupling procedures similar to those which properly describe the standard Anderson model
within the Kondo regime and outside of it. For an arbitrary ratio $x$ between the amplitudes of correlated  
and single-particle hopping terms, the transport GF fulfils the  $x \leftrightarrow 2-x$ 
symmetry of the model. The average occupation of the dot also obeys this symmetry, albeit  
the spectral function of the quantum dot, calculated within an analogous decoupling scheme
as for the transport GF, does not. We identify the physical reason for this behavior, and 
propose a way to cure it. Since the correlated-hopping term breaks the particle-hole symmetry of the model 
and  modifies all transport characteristics of the system, the detailed knowledge of its 
influence on measurable characteristics is a prerequisite for its experimental detection. 
Simple, experimentally feasible methods are proposed.  
\end{abstract}
\maketitle

\section{Introduction}
The study of quantum dots~\cite{alivisatos1996} continues to enjoy a high popularity. 
Transport properties of nanostructures consisting of quantum dots (QDs) or molecules placed 
between two or more external electrodes have been intensively investigated in the past few 
decades~\cite{benenti2017,zimbovskaya2011}, with the goal, {\it e.g.}, to achieve   
efficient heat to electricity conversion at the nanoscale. Nanodevices with ferromagnetic or/and 
superconducting leads may be relevant as sources of a pure spin current \cite{purespin} or  
entangled electrons \cite{entangled}, needed for spintronics and quantum information technology~\cite{qi}.
For example, spin valves have a high potential for controlling spin
currents~\cite{spinvalves}.

The properties and functionalities of such structures strongly depend on the state of the leads, 
their coupling to the central region~\cite{diventra2008}, the interactions of the electrons on the central region, 
and on external conditions like temperature, magnetic field, {\it etc.} The experimental control
of the relevant parameters and the theoretical understanding of their effect on measurable characteristics of 
devices is at the heart of their application potential. 

The standard theoretical modelling of such systems is based on the Anderson Hamiltonian:
\be
H=H_{LR}+H_C+H_{\mathrm{tunn}},
\ee
where $H_{LR}$ describes the external leads ($L$: left lead, $R$: right lead), $H_C$ the central region 
typically containing the Hubbard \cite{hubbard1963} repulsion, 
and $H_{\mathrm{tunn}}$ the coupling between the leads and the central QD. 
The coupling is visualised as the tunneling of electrons between electrodes 
and QD, and is described by the following term in the Hamiltonian: 
\be
H_{\mathrm{tunn}}=\sum_{\lambda k \sigma}\left(V_{\lambda k \sigma}c^\dagger_{\lambda k \sigma}d_\sigma + \mathrm{h.c.}\right).
\label{eq:single-p}
\ee
The amplitude $V_{\lambda k \sigma}$ is a single-particle
transfer proportional to $\langle \Psi_{\lambda k\sigma}|h(\mathbf{r})|\Psi_{c \sigma} \rangle$, 
where $\Psi_{c \sigma}$ and $\Psi_{\lambda k \sigma}$ are the wave functions of the central region 
and the extended states in the leads ($\lambda = L,R$), respectively, 
and $h(\mathbf{r})$ the single-particle part 
of the first-quantised Hamiltonian of the system. It turns out that besides the single-particle term
there may exist another term promoting the transfer of electrons from the electrodes to the central 
region and {\it vice versa}. This term, with the amplitude denoted by $K_{\lambda k \sigma}$, 
has a many-body origin, with 
\be
K_{\lambda k \sigma} =
\langle\langle\Psi_{c \bar{\sigma}}\Psi_{\lambda k \sigma}|V_{C}(\mathbf{r}-\mathbf{r}^\prime)|\Psi_{c \sigma}\Psi_{c \bar{\sigma}}\rangle\rangle .
\ee 
In second-quantised representation, the corresponding Hamiltonian is given by
\be
H_{\mathrm{ass}}=\sum_{\lambda k \sigma}\left(K_{\lambda k \sigma}c^\dagger_{\lambda k \sigma}d_\sigma n_{\bar{\sigma}} + \mathrm{h.c.}\right),
\label{eq:corr-h}
\ee
where  ${\bar{\sigma}}$ denotes spin opposite to $\sigma$, {\it i.e.}, ${\bar{\sigma}} = -\sigma$. 
$H_{\mathrm{ass}}$ is known as assisted or correlated hopping~\cite{micnas1989}. 
Due to its dependence on the charge state of the central site, it is also called charge-bond 
interaction~\cite{hubsch2006}. Apparently, 
this term describes the transfer of a spin-$\sigma$ electron between the dot and the electrode, 
provided another electron with opposite spin occupies the dot. 

Both terms have been of considerable interest in studies of strongly correlated bulk materials. 
The single-particle hopping $V$  has been intensively investigated in the context of heavy 
fermions \cite{wysokinski2016}, where it describes 
the coupling between, {\it e.g.}, $d$ and $f$ orbitals. On the other hand, the presence of the correlated-hopping term $K$ 
in solids has been found to affect collective properties of materials; it was mainly studied in the context
of high-temperature superconducting cuprates as the interaction promoting the appearance of the superconducting 
instability~\cite{micnas1991}, and/or explaining the asymmetry between the superconducting domes of electron-doped 
and hole-doped compounds~\cite{wysokinskimm2017,zegrodnik2017}.  

In the context of nanodevices, correlated hopping is expected to be present in most cases. However,
this term has not attracted the attention it probably (in our opinion) deserves.
In fact, studies of correlated hopping in the context of transport {\it via} nanostructures 
are sparse~\cite{meir2002,guinea2003,borda2004,stauber2004,lin2007,tooski2014,gorski2019}, and 
mainly by numerical techniques.
Correlated hopping has been proposed to explain anomalous features
observed {\it inter alia} in transport through quantum point contacts \cite{cronenwett2002} and 
single-electron molecular transistors \cite{yu2005}, but---as far as we know---no compelling evidence 
exists on its experimental relevance. This calls for detailed theoretical studies, in order to find 
and quantify possible ways for its experimental detection.

In this work, we present a systematic analysis of the role of correlated hopping 
in transport {\it via} QDs, {\it i.e.}, the modifications it introduces to the standard 
behavior of the single-impurity Anderson model with Hubbard-only interaction. To solve the  
problem analytically, and to gain a deeper insight into the physics of correlated hopping, the equation 
of motion (EOM) technique is employed. We return to this aspect in the concluding section.

While the EOM method does not provide the exact GFs 
(except for non-interacting systems), as it relies 
on decouplings and projections of higher-order GFs onto lower order ones, it is 
easy to implement in quite arbitrary situations and for (almost) arbitrary Hamiltonians,
in the linear (small voltage) regime and beyond.
Nevertheless, it has to be noted that the near-equilibrium results of~\cite{lin2007,tooski2014} are based
on the numerical renormalization group (NRG) technique, which is known to capture correlation effects
in an essentially exact manner, as discussed recently in some detail~\cite{costi2019}.

The goals of the paper are:
(i) to generalise the EOM method, earlier applied to the Anderson Hamiltonian (where it properly describes 
Kondo correlations \cite{lavagna2015,eckern2020}), to the model with correlated hopping; 
(ii) to analyze the role of this contribution, which breaks particle-hole symmetry, on the Kondo peak (its width, 
temperature dependence, etc.) and the transport characteristics of a two-terminal system;
and (iii) to determine kinetic and transport coefficients of the two-terminal QD, and
identify experimental signatures of the extra term.
Last but not least, we demonstrate the power of the EOM approach by presenting
selected results beyond the linear regime.

In the next section, Sec.~\ref{sec:model}, we describe the model
and its parametrization used throughout the paper, and express the charge and heat currents
in terms of the appropriate GF. The full set of equations for the transport GF
is introduced in Sec.~\ref{sec:GFs}. The results are presented and discussed in 
Sec.~\ref{sec:res} and Sec.~\ref{sec:transport}. The summary and conclusions are given  
in Sec.~\ref{sec:summconcl}. Some technical details and lengthy calculations are relegated 
to the appendices. The Supplementary Material~\cite{sm} contains additional details.

\section{The model and basic definitions}
\label{sec:model}
The Hamiltonian to be studied is similar to the standard single-impurity Anderson model, albeit 
modified to include correlated hopping:
\beq
{H}&=&\sum_{\lambda {k} \sigma}\varepsilon_{\lambda {k}}n_{\lambda {k} \sigma} + 
\sum _{\sigma} \varepsilon _{\sigma}n_{\sigma} +Un_\uparrow n_\downarrow \nonumber \\
&+&\sum_{\lambda {k} \sigma} \left({V}_{\lambda {k}\sigma} c^{\dagger}_{\lambda k\sigma} D_{\sigma} 
+ {V}_{\lambda {k}\sigma}^* D^{\dagger}_{\sigma} c_{\lambda {k} \sigma}\right),
\label{eq:ham1}
\eeq
where $n_{\lambda {k} \sigma}=c^{\dagger}_{\lambda {k} \sigma}c_{\lambda {k} \sigma}$~
and $n_{\sigma}=d^{\dagger}_{\sigma} d_{\sigma}$
denote particle number operators for the leads and the dot, respectively.  
The operators $c^{\dagger}_{\lambda k\sigma} (d^{\dagger}_{\sigma})$ create electrons in
respective states $\lambda {k}\sigma$ $(\sigma)$ in the lead $\lambda$ (on the dot). 
The energies of the leads are measured from their chemical potentials $\mu_\lambda$,
so $\varepsilon_{\lambda k}=\varepsilon_{0\lambda k}-\mu_\lambda$, with the dependence
of $\varepsilon_{0\lambda k}$ on $\lambda$ allowing for a different spectrum in each of the leads. 
The spin is $\sigma=\pm 1$ $(\uparrow, \downarrow)$, and $\varepsilon_\sigma=\varepsilon_d+\sigma \mu_B B$, 
with $B$ the magnetic field, $\mu_B$ the Bohr magneton, 
and $\varepsilon_d$ the dot electron energy level.
The Hubbard parameter $U$ describes the repulsion between two electrons on the dot. 
The operator $D_\sigma=d_\sigma(1-xn_{\bar{\sigma}})$ takes care of the occupation dependence of the hopping.

The state-dependent hopping has been parameterised by $x$ which is minus the ratio between $K_{\lambda k \sigma}$ and 
${V}_{\lambda k\sigma}$,
$x= - {K_{\lambda k \sigma}}/{V_{\lambda k \sigma}}$. 
The parameter $x$, in principle, may be complex, and even in case it is real it may
have both positive and negative values \cite{lin2007,tooski2014}. 
For simplicity, it is assumed not to depend on $\lambda k \sigma$, 
{\it i.e.}, to have the same constant, spin and wave vector independent value for both leads.
This assumption should hold provided the two leads are composed of similar (or even identical)
materials. Generally we also expect the $k$ dependence to be of lesser importance, as usual in
Fermi liquid theory. However, the spin dependence may become relevant for magnetic leads.
Here, following \cite{lin2007,tooski2014}, we assume $x$ to be real, and focus on the 
interval $0 \le x \le 2$.

\subsection{Currents in the two-terminal system}
The charge current and energy current flowing out of the electrode $\lambda$ are calculated
as the time derivative of the average charge, $\langle N_{\lambda}\rangle=\sum_{k\sigma} \langle n_{\lambda{k}\sigma}\rangle$, respectively average energy $\langle H_\lambda \rangle =\sum_{{k}\sigma}\varepsilon_{\lambda {k} \sigma}\langle n_{\lambda {k} \sigma}\rangle$ of lead $\lambda$. 
The derivation is sketched in App.~\ref{app:curr}.
Application of those results to the two-terminal QD we are interested in here,
provides  $I=I_L=-I_R$, which expresses current conservation in the system: 
\be
I=\frac{2e}{\hbar} \sum_\sigma \tilde{\Gamma}_\sigma\int\frac{dE}{2\pi}\left[f_L(E)-f_R(E)\right]\mathrm{Im} G^r_\sigma(E), 
\label{eq:c-curr-wbl}
\ee
with $\tilde{\Gamma}_\sigma={\Gamma_\sigma^L\Gamma_\sigma^R}/\left({\Gamma_\sigma^L+\Gamma_\sigma^R}\right)$; the parameters
\be
\Gamma_\sigma^{\lambda}(E)=2\pi\sum_{{k}}|V_{\lambda{k}\sigma}|^2\delta(E-\varepsilon_{\lambda{k}})
\ee
describing the coupling between the dot and the electrode are assumed to be independent of energy $E$, which corresponds to the wide-band limit. Similar expressions can be derived 
for the heat current flowing from the left,  
\be
J_L=\frac{2e}{\hbar} \sum_\sigma \tilde{\Gamma}_\sigma \int\frac{dE}{2\pi} (E-\mu_L)\left[f_L(E)-f_R(E)\right] \mathrm{Im} G^r_\sigma(E),
\label{h-curr-wbl}
\ee
and right electrodes:
\be
J_R=\frac{2e}{\hbar} \sum_\sigma \tilde{\Gamma}_\sigma \int\frac{dE}{2\pi} (E-\mu_R)\left[f_R(E)-f_L(E)\right] \mathrm{Im} G^r_\sigma(E).
\ee
It can also be verified that
\be
\dot{Q}+(\mu_L-\mu_R)I=0
\label{en-cons}
\ee
in agreement with energy conservation: here $\dot{Q}=J_L+J_R$ is the total heat current leaving the leads. 

We emphasize that the transport GF, 
\be
G^r_\sigma(E)=\langle\langle  D_\sigma|D^\dagger_\sigma\rangle\rangle^r_{E},
\label{eq:gf-transport}
\ee
differs from the spectral one, 
\be
g^r_\sigma(E)=\langle\langle  d_\sigma|d^\dagger_\sigma\rangle\rangle^r_{E},
\label{eq:gf-spectral}
\ee
the latter {\it inter alia} describing the occupation $\langle n_\sigma\rangle$ of the quantum dot. The function 
$$
f_{L/R}(E)= \left[ \exp\frac{(E-\mu_{L/R})}{k_BT_{L/R}}+1 \right]^{-1}
$$
is the Fermi distribution function describing the electrons in the lead $L/R$, assumed to be in equilibrium at temperature $T_{L/R}$ and chemical potential $\mu_{L/R}$.

The above formulae for the currents are valid for arbitrary voltages $V=(\mu_R-\mu_L)/e$, where $e$ is the electron 
charge. In particular, Eq.~(\ref{eq:c-curr-wbl}) allows the calculation of the conductance beyond the linear regime.
In the general ($V \neq 0$) case, we define the differential conductance as
\be
G_d (V)=\frac{\partial I(V)}{\partial V}.
\ee 

\subsection{Linear transport coefficients} 
Assuming the temperature difference between the right and left electrode $\Delta T=T_R-T_L$ as
well as the voltage $V$ to be small parameters, we can expand the formulae (\ref{eq:c-curr-wbl}) 
for the current across the system and the similar one for the heat flux $\dot{Q}=J_L+J_R$, in order to
obtain the (symmetric) Onsager matrix of linear kinetic coefficients $L_{ij}$, as well as the related set of transport 
parameters: the conductance ($G$), the Seebeck coefficient ($S$), and the thermal conductance
($\kappa$)~\cite{mahan,zlatic2014,benenti2017}. 
In the present model, 
the linear coefficients are given by the moments $M_n$ of the imaginary part of the transport GF,
\be
M_n(T)=\int dE \left(- f^\prime \right)
(E-\mu)^n\sum_{\sigma}\tilde{\Gamma}_\sigma
\left(\frac{-1}{\pi}\right) \mathrm{Im} G^r_\sigma(E)
\label{moments}
\ee
where $f^\prime = {\partial f(E,T)}/{\partial E}$; here
we set $\mu =\mu_L =\mu_R$, and $T=T_L=T_R$. The linear conductance and Seebeck coefficient read 
\be
G=\frac{2e^2}{h}M_0(T),
\label{przew-lin}
\ee

\be
S=\frac{k_B}{e}\frac{1}{k_BT}\frac{M_1(T)}{M_0(T)}.
\label{tep-lin}
\ee

It has to be noted that the transport density of states, 
$N_{\mathrm{tr}}(E)=({-1}/{\pi})\mathrm{Im}\langle \langle  D_\sigma|D^\dagger_\sigma\rangle\rangle^r_{E}$,
may have features narrow on the scale of $k_BT$. In such a case the Sommerfeld low-temperature 
expansion \cite{ashcroft-mermin} 
is not valid, and one has to use the above expressions to calculate the linear Seebeck coefficient. 
For parameters such that $N_{\mathrm{tr}}(E)$ is a smooth function of energy on the scale $k_BT$ around the chemical potential, 
one finds the approximate formulae
\beq 
G\approx\frac{2e^2}{h}N_{\mathrm{tr}}(\mu), 
\label{przew-lin-approx} \\
S\approx \frac{\pi^2}{3}\frac{k_B}{e}\frac{N_{\mathrm{tr}}'(\mu)}{N_{\mathrm{tr}}(\mu)},
\label{mott-s}
\eeq
where the prime means the derivative with respect to energy.

\section{Calculation of the Green functions}
\label{sec:GFs}
To calculate the spin-dependent retarded GFs~\cite{haug-jauho1996} $G^{r}_\sigma(\omega)$ 
and $g^{r}_\sigma(\omega)$, we use the EOM method~\cite{zubarev1960} and the approximation scheme
known as Lacroix approximation~\cite{theuman1969,lacroix1981,lacroix1982},
with some important extensions proposed recently~\cite{lavagna2015}. In this section and the following ones, we shall (mostly)
work in units such that $\hbar=k_B =1$. 
We also use frequency as argument of all GFs below, and omit the $r$ subscript with the understanding that 
we shall first calculate the retarded GFs, and advanced and lesser GFs will be obtained from them by known relations \cite{sm}.

\begin{widetext}
\subsection{Transport Green function}
The transport GF $\langle\langle D_{\sigma} |D^{\dagger}_{\sigma}\rangle\rangle _{\omega}$ has been calculated in App.~\ref{app:trGF}, 
and we only quote the final formula here:
\beq
\langle\langle D_\sigma|D^\dagger_\sigma\rangle\rangle_\omega=\frac{1-x(2-x)(\langle n_{\bar{\sigma}}\rangle+\tilde{b}_{1\bar{\sigma}})+n^D_{\mathrm{eff}}(\omega)I_D(\omega)}{\omega-\varepsilon_d-\Sigma_{0\sigma}+\Sigma_D(\omega)},
\label{a-sol-gf-t}
\eeq
where
\be
I_D(\omega)=\frac{U-x(2-x)(\Sigma_{0\sigma}+\Sigma^{(1)}_{\bar{\sigma}})}{\omega-\varepsilon_\sigma-U-\Sigma_{ID}(\omega)},
\label{a-sol-sp2-t}
\ee
and
\beq
n^D_{\mathrm{eff}}(\omega)&=&(1-x)^2(\langle n_{\bar{\sigma}}\rangle+\tilde{b}_{1\bar{\sigma}})-\bar{b}_{2\bar{\sigma}}, \\
B_D(\omega)&=&(1-x)^2[\tilde{b}_{1\bar{\sigma}}\Sigma_{0\sigma} -\Sigma^T_{1\bar{\sigma}}-\Sigma^T_{2\bar{\sigma}}]
- \bar{b}_{2\bar{\sigma}}\Sigma_{0\sigma}, \\
\Sigma_D(\omega)&=&x(2-x)(\tilde{b}_{1\bar{\sigma}}\Sigma_{0\sigma}-
\Sigma^T_{1\bar{\sigma}})-I_D(\omega)B_D(\omega), \\
\Sigma_{ID}(\omega)&=&(1-x)^2(\Sigma_{0\sigma}+\Sigma^{(1)}_{\bar{\sigma}})
- x(2-x)\Sigma^T_{2\bar{\sigma}}+\Sigma_{\bar{\sigma}}^{(2)}.
\label{a-sol-sp2-tD}
\eeq
For various definitions and details, see App.~\ref{app:trGF}, in particular, Eq.~(\ref{sol-gf-t}) in 
conjunction with Eqs.~(\ref{sol-sp2-t}) and (\ref{sol-b44})--(\ref{sol-sp2-tD}).

\subsection{Spectral Green function}
The  following expression for the spectral GF $\langle\langle d_{\sigma} |d^{\dagger}_{\sigma}\rangle\rangle _{\omega}$ has been obtained 
by employing the same decoupling scheme as above \cite{sm}:
\beq
\langle\langle d_\sigma|d^\dagger_\sigma\rangle\rangle_\omega=\frac{1-x(\tilde{b}_{1\bar{\sigma}}+\tilde{b}_{2\bar{\sigma}})+n^d_{\mathrm{eff}}(\omega)I_d(\omega)}{\omega-\varepsilon_d-\Sigma_{0\sigma}
+x[(\tilde{b}_{1\bar{\sigma}}+\tilde{b}_{2\bar{\sigma}})\Sigma_{0\sigma}-
\Sigma^T_{1\bar{\sigma}}+(1-x)\Sigma^T_{2\bar{\sigma}}]-I_d(\omega)B_d(\omega)},
\label{sol-gfd}
\eeq
where 
\be
I_d(\omega)=\frac{U-x(2-x)\Sigma_{0\sigma}+x(1-x)\Sigma^{(1)}_{\bar{\sigma}}-x\Sigma^{(2)}_{\bar{\sigma}}
-x^2[\Sigma^T_{1\bar{\sigma}}+\Sigma^T_{2\bar{\sigma}}
-(\tilde{b}_{1\bar{\sigma}}+\tilde{b}_{2\bar{\sigma}})\Sigma_{0\sigma}]}{\omega-\varepsilon_\sigma-U-(1-x)^2(\Sigma_{0\sigma}+\Sigma^{(1)}_{\bar{\sigma}})-\Sigma_{\bar{\sigma}}^{(2)}-x(\tilde{b}_{2\bar{\sigma}}\Sigma_{0\sigma}-\Sigma^T_{2\bar{\sigma}})+x(1-x)(\tilde{b}_{1\bar{\sigma}}\Sigma_{0\sigma}-\Sigma^T_{1\bar{\sigma}})},
\label{sol-id}
\ee
and 
\beq
n^d_{\mathrm{eff}}(\omega)&=&\langle n_{\bar{\sigma}}\rangle+(1-x)\tilde{b}_{1{\sigma}}-\bar{b}_{2{\sigma}} \\
B_d(\omega)&=&[(1-x)\tilde{b}_{1\bar{\sigma}}- \bar{b}_{2\bar{\sigma}}]\Sigma_{0\sigma} 
-(1-x)(\Sigma^T_{1\bar{\sigma}}+\Sigma^T_{2\bar{\sigma}}). 
\eeq
\end{widetext}
It has to be stressed again that both GFs, {\it i.e.}, the spectral and transport one are 
coupled together. They both have to be calculated simultaneously as various quantities they depend on 
require the knowledge of both of them. Needless to say that for $x=0$ the transport GF (\ref{a-sol-gf-t}) reduces to the 
spectral one (\ref{sol-gfd}) as it has to be, and that the result 
agrees with the formula found earlier by Lavagna \cite{lavagna2015}.

The symmetry of the Hamiltonian suggests that both GFs are symmetric with respect to $x=1$. The approximate transport GF given in Eq.~(\ref{a-sol-gf-t}) is indeed symmetric, and calculated for $x=0$ even analytically is the same as for $x=2$.  However, this is not correct for the spectral function, Eq.~(\ref{sol-gfd}): its value at $x=2$ is not the same as that for $x=0$. This issue, which, however, does not affect the symmetry of the transport coefficients, will be discussed later on.

In a recent paper \cite{eckern2020} the GF for the $x=0$ model has been studied in the context of
a three-terminal QD, which in the strongly non-equilibrium limit works as a heat engine. We have shown that  
the above formula is quantitatively correct in describing the spectral and transport properties 
of the QD in the Kondo regime, even in the particle-hole-symmetric case which is notoriously difficult to capture by 
the EOM technique. This is true in equilibrium as well as far from it.
In Sec.~\ref{sec:res} we shall discuss the transport characteristics of
the {\em two-terminal} quantum dot with correlated hopping, but with focus on the linear regime.

\subsection{Lifetimes: second order calculations}
When writing the expressions for various self-energies, we 
have introduced the parameters $\tilde{\gamma}^{\sigma}_{1}$ and $\tilde{\gamma}_{2}$, 
which replace the infinitesimal parts $\gamma=0^+$ in the self-energies; clearly, they represent the inverse lifetimes 
of singly and doubly occupied states on the dot, respectively. These decay rates take into
account higher-order processes neglected at the present level of approximation. The importance of including such
decay rates for the proper description of the Kondo resonance has been observed 
in~\cite{vanroermund2010}, and found to result from higher-order processes. Lavagna argued later~\cite{lavagna2015} 
that they can be calculated perturbatively using Fermi's golden rule. She also noted that fourth order 
contributions vanish for systems in equilibrium and without magnetic field.
\begin{figure}
\includegraphics[width=0.85\linewidth]{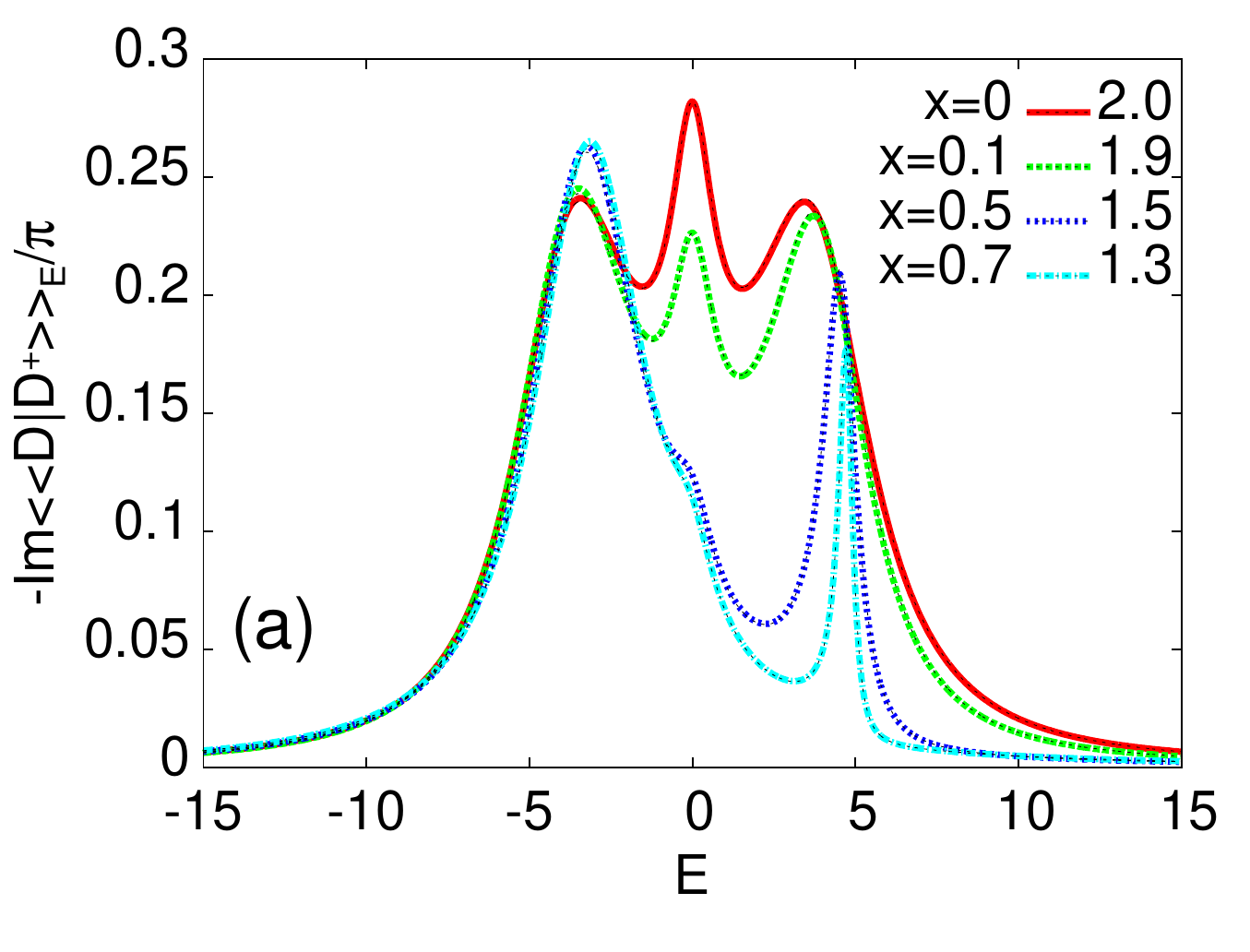}
\includegraphics[width=0.85\linewidth]{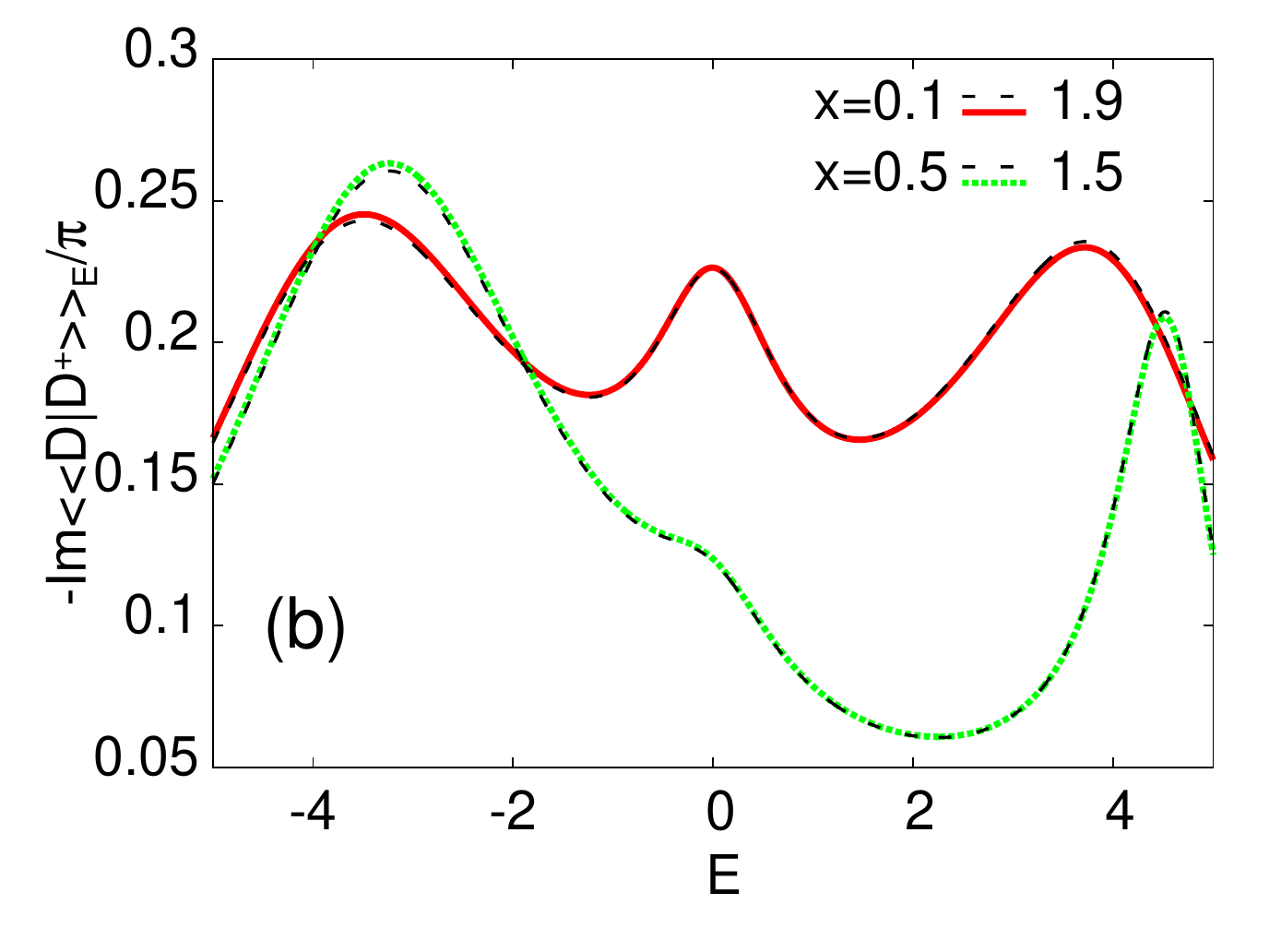}
\caption{(color online)
The transport GF of the model at hand is found to be symmetric with respect to $x=1$
to a very good approximation.
In panel (a) we show the symmetry for the imaginary part of the transport GF,
calculated for a few values of $x$ (as indicated) and $2-x$ (overlapping dashed curves). 
The differences between the two curves at particular values
of the energy are smaller than $1\%$. Panel (b) shows the detailed behavior 
for $x=0.1$ {\it vs.} 1.9, and $x=0.5$ {\it vs.} 1.5, in the region where the differences are largest.
The other parameters are $\varepsilon_d=-4$, $U=8$, and $T=0.3$.}
\label{fig:rys1-1}
\end{figure} 

The direct usage of Fermi's golden rule,
\be
\tilde{\gamma}_i=2\pi\sum_{f}|\langle f|V_I|i\rangle |^2 \delta(E_i-E_f) ,
\ee
with $V_I$ being the tunneling part of the Hamiltonian (\ref{eq:ham1}), leads
to the following expressions valid to second order in the dot coupling $V_{\lambda k \sigma}$:
\beq
\tilde{\gamma}^\sigma_1&=&\sum_{\lambda}\left(\Gamma^\lambda_\sigma [1-f_\lambda (\varepsilon_\sigma)]
+(1-x)^2\Gamma^\lambda_{{\sigma}} f_\lambda (\varepsilon_{{\sigma}}+U)\right)
\nonumber \\ \\
\tilde{\gamma}_2&=&(1-x)^2\sum_{\lambda \sigma}\Gamma^\lambda_\sigma [1-f_\lambda (\varepsilon_\sigma+U)].
\eeq
The last equation shows that the contribution from the doubly occupied 
states vanishes for $x=1$, as  expected: for this value of $x$
the doubly occupied state is totally decoupled from the system.

\section{Spectral and transport GFs: Numerical results and symmetry discussion}
\label{sec:res}
This section is devoted to the presentation of the results for the system in equilibrium, and for 
vanishing external magnetic field and spin-independent tunnelings. The transport coefficients 
can be calculated in the linear regime {\it via} Eqs.~(\ref{moments})--(\ref{tep-lin}). In the following, 
all energies are measured in units of $\Gamma_0=\Gamma^L_\downarrow=\Gamma^L_\uparrow$. 
We start the discussion with the imaginary parts of the transport and spectral GFs, {\it i.e.},
the transport and dot's densities of states. 

\subsection{The Green functions and their symmetry}
\begin{figure}
\includegraphics[width=0.85\linewidth]{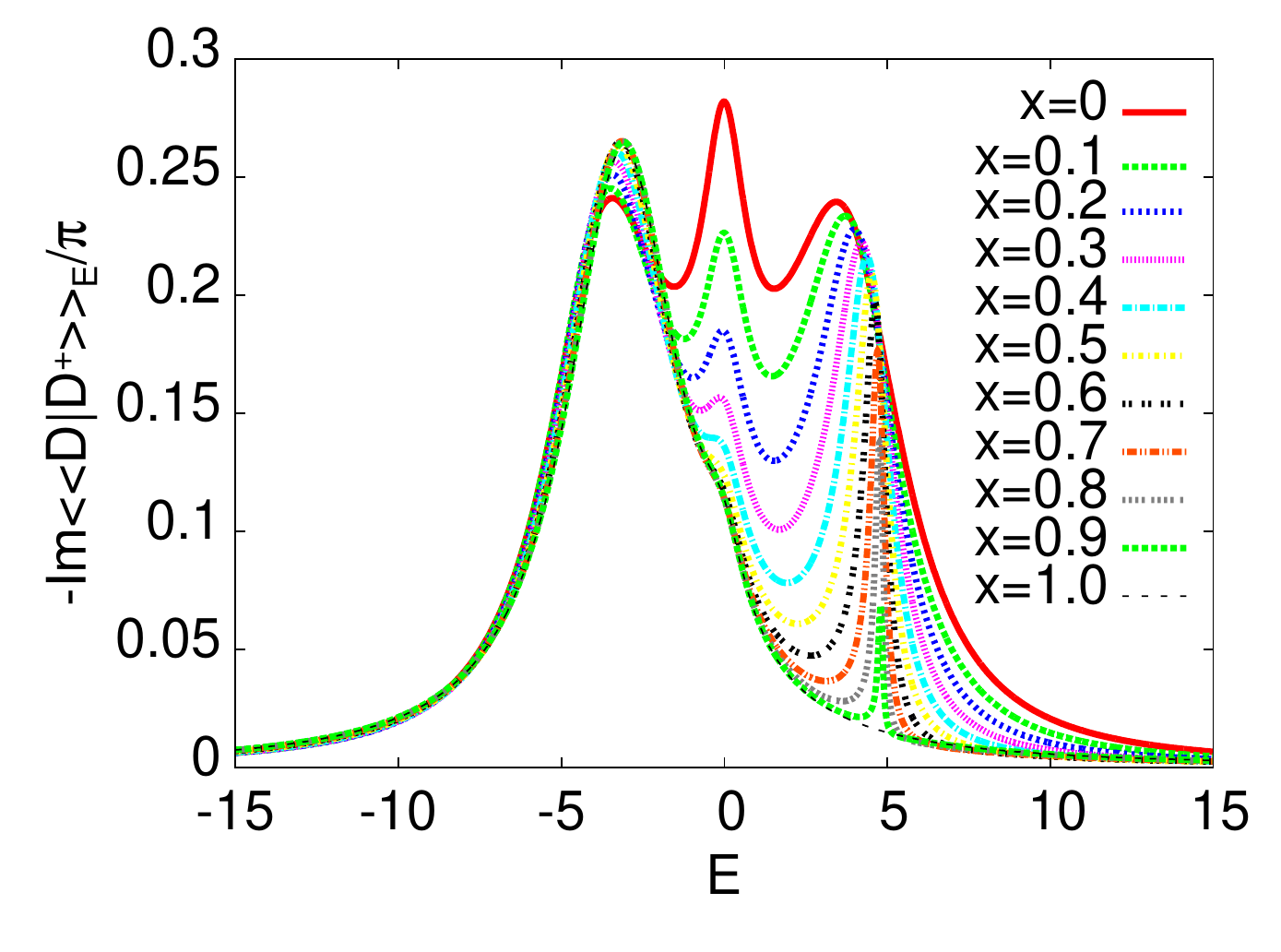}
\caption{(color online) The imaginary part of the transport GF  
{\it vs.} energy $E$ for a number of $x$ values. For $x=1$ the doubly occupied state is blocked.
This is visible as the disappearance of the upper Hubbard band. The other parameters for this 
particle-hole symmetric model read $\varepsilon_d=-4$, $U=8$, and $T=0.3$.}
\label{fig:rys2}
\end{figure}

\begin{figure}
\includegraphics[width=0.85\linewidth]{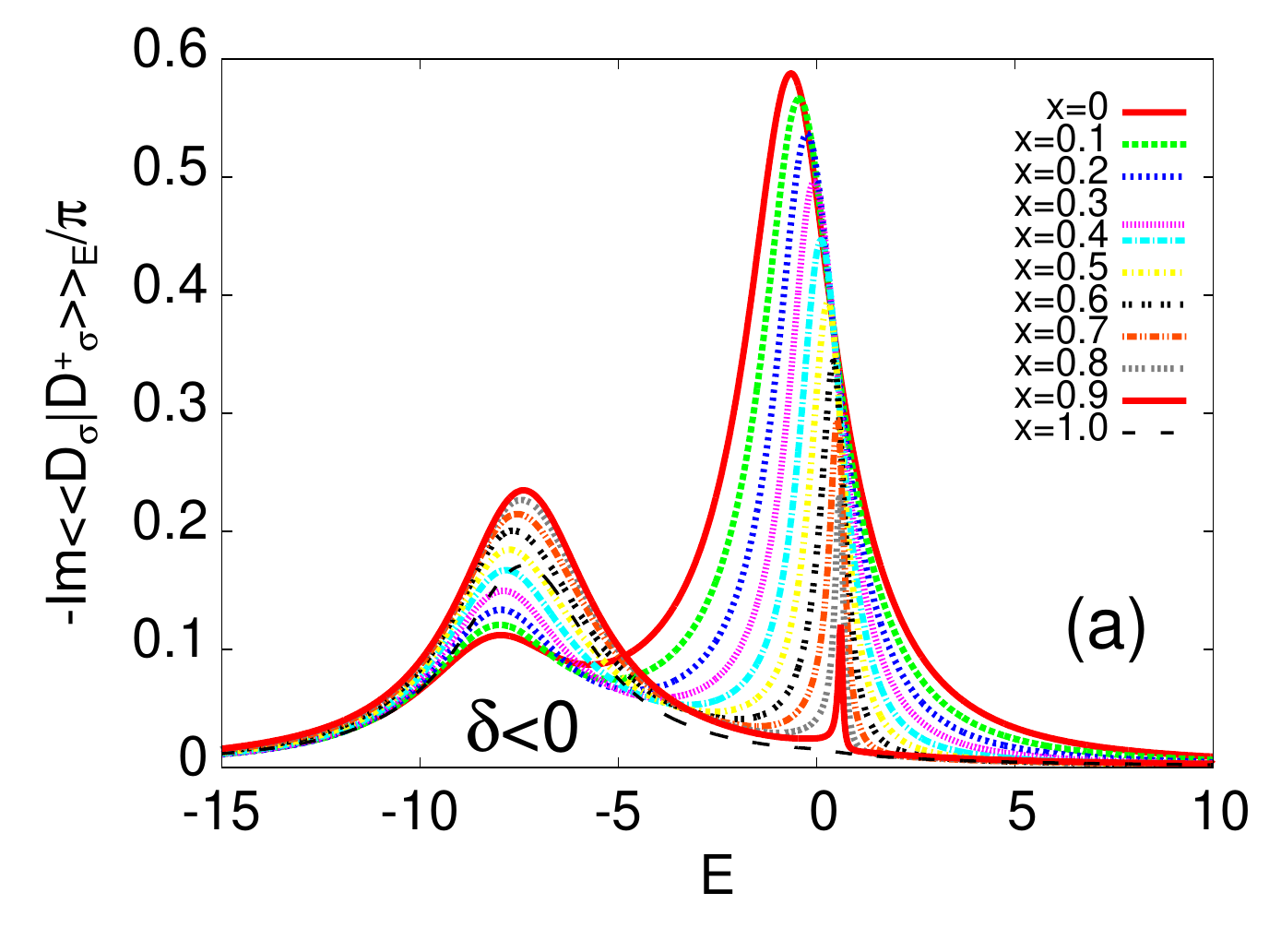}
\includegraphics[width=0.85\linewidth]{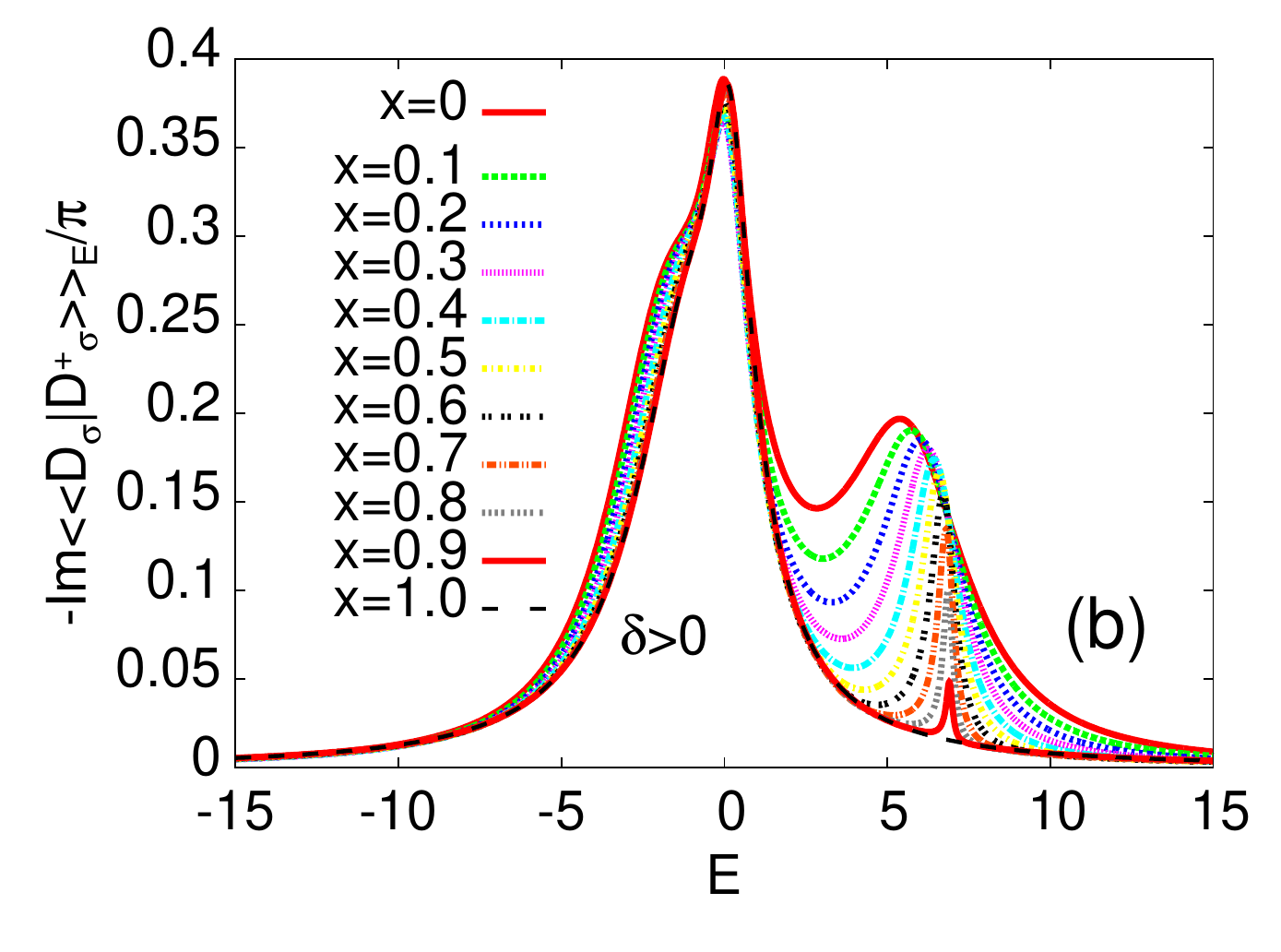}
\caption{(color online) Panel (a) shows the transport density of states as a function
of energy $E$ calculated for a set of $x$ values for negative $\delta=\varepsilon_d+U/2$, 
with $\varepsilon_d=-8$, $U=8$, and $T=0.3$. Panel (b) shows the similar evolution for $\delta=2>0$ and 
$\varepsilon_d=-2$.}
\label{fig:rys6a} 
\end{figure}

\begin{figure}
\includegraphics[width=0.9\linewidth]{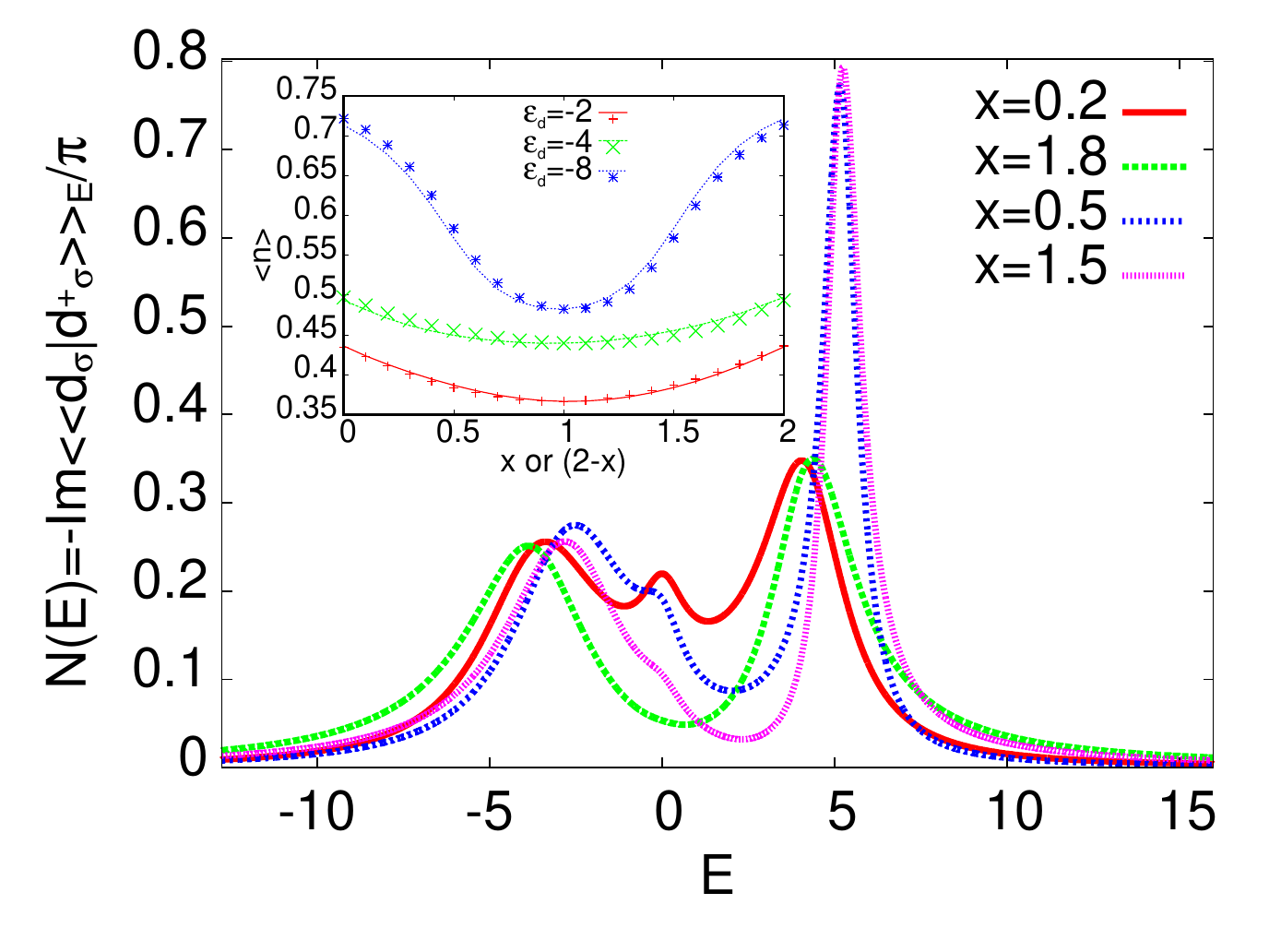}
\caption{(color online) The dot's spectral density of states,
$N(E)=-({1}/{\pi}) \mathrm{Im}\langle\langle d_\sigma|d^\dagger_\sigma\rangle\rangle_E$, {\it vs.} energy for
several $x$ values. $N(E)$ changes with $x$ but is not symmetric, $N(E,x)\ne N(E,2-x)$. 
The other parameters are $\varepsilon_d=-4$, $U=8$, and $T=0.3$.
The inset shows
the $x$ dependence of the average occupation $\langle n\rangle$ per spin for $\varepsilon_d=-2,-4,-8$. 
For this quantity, the departures from perfect symmetry are very small. In order to increase their
visibility, we plot the dependence
$\langle n\rangle$ {\it vs.} $x$ as solid lines, and the dependence $\langle n\rangle$ {\it vs.} 
$(2-x)$ as points of the same color.}
\label{fig:rys3}
\end{figure}

A brief inspection of the formulae (\ref{a-sol-gf-t}) and (\ref{sol-gfd}) for the transport and 
spectral GFs suggests that the former is symmetric with respect to changes of $x$ by $(2-x)$, 
while the latter is  more difficult to judge due to a more complicated $x$ dependence. Thus we resort 
to numerical calculations and postpone further discussion of these formulae to the end
of this section.
In Figs.~(\ref{fig:rys1-1})--(\ref{fig:rys4}) we show the imaginary 
parts of both GFs as a function of energy for a number of $x$ values. The numerically perfect symmetry 
of the transport GF with respect to $x=1$ is visible in Fig.~(\ref{fig:rys1-1}a). The imaginary part of
the transport GF is plotted as function of energy for a number of $x$ and symmetry-related $(2-x)$ values. 
The (hardly visible) dashed black curves for the parameter $(2-x)$ overlap with the colored curves for $x$.
Panel (b) of the figure is a magnification of the  
part of panel (a) close to the maxima, where the differences are largest. The observed 
differences are typically smaller than $1\%$.

The systematic evolution of the transport density of states  
with changing $x$ is shown in Fig.~(\ref{fig:rys2}). We present the results 
for $x\in [0,1]$ only, as the curves for the other values of $x$ are related by symmetry. 
For illustration we have 
assumed a particle-hole symmetric situation with $\varepsilon_d=-4$, $U=8$, and the same temperature
for both leads, $T=0.3$. For these parameters and $x=0$, both transport and spectral GFs are identical, 
and the quantum dot is in the Kondo regime. Hence one observes the Abrikosov-Suhl resonance, also known
as Kondo resonance, at the chemical potential $\mu=0$, and two Hubbard bands located symmetrically 
around zero energy.  The occurrence of the Kondo effect in the particle-hole symmetric Hubbard 
model shows the power of the present version of the EOM technique supplemented with lifetime 
effects~\cite{lavagna2015}. 

The increase of $x$ results in distinctive modifications of the transport 
density of states. First, one notices that the curves (for $x\ne 0$) are no longer particle-hole symmetric.
Concomitant with this observation is the modification of the 
Kondo resonance, which develops a strong asymmetry with respect to the chemical potential ($E=0$). The lower Hubbard
band changes rather weakly with $x$. Its height slightly increases, and the position slightly moves towards the 
chemical potential. Most dramatic changes are apparent in the upper Hubbard band, which strongly
decreases with increasing $x$, getting narrower and finally vanishing completely for $x=1$. For the considered parameters the center 
of the upper Hubbard band moves slightly to the right in the figure. 
The result for  $x=1$ requires additional comments. Let us note
that the operator $D_\sigma$ for the doubly occupied  state with $n_{\bar{\sigma}}=1$ vanishes. Under these 
conditions, the upper Hubbard band composed of the doubly occupied states does not contribute to transport.

In Figs.~(\ref{fig:rys1-1}) and (\ref{fig:rys2}) we have shown the evolution of the transport 
density of states with $x$ for the particle-hole symmetric model for which $\delta=\varepsilon_d+U/2=0$. 
Non-zero values of $x$ break the particle-hole symmetry of the model. 
It turns out that for arbitrary values of $\delta$ the transport GF is symmetric 
with respect to $x$, albeit the differences are slightly larger than those in Fig.~(\ref{fig:rys1-1}). 
Moreover, the effect of $x$ on the transport density of states 
varies depending on whether $\delta$ is positive or negative. In Fig.~(\ref{fig:rys6a}) we illustrate 
this for $U=8$ and two values of $\varepsilon_d$. 
In panel (a) of the figure we choose $\varepsilon_d=-8$, leading to negative value of $\delta=-4$.
One can see that for this set of parameters the transport density of states $N_{\mathrm{tr}}(\mu)$ 
at the chemical potential ($\mu=0$) strongly changes with $x$. Both the absolute value and 
the slope are affected. Considering the formulae (\ref{przew-lin}) and (\ref{tep-lin}) for the 
transport parameters, which strongly depend on the transport density of states close to $E=0$, 
one expects for these parameters a noticeable changes of both conductance and thermopower with $x$.
On the other hand, for the set of parameters used in the Fig.~(\ref{fig:rys6a}b), the transport GF 
for energies close to the chemical potential hardly changes with $x$;
thus both conductance and thermopower are expected to vary only slightly with $x$.  
The symmetry of the transport GF with respect to $x\leftrightarrow 2-x$ ensures, 
as we shall see in the next section, the same symmetry of the transport coefficients.

\subsection{The \textit{x~vs.~(2-x)} symmetry in further detail}
\label{sec:x2mx}
As argued in detail in \cite{{tooski2014}}, the model at hand is symmetric under the transformation
$x \leftrightarrow 2-x$.
However, we find that the (approximate) spectral GF derived above does not obey this symmetry, in contrast to the
(approximate) transport GF. In particular, this deficiency is already apparent in the analytical formula for the
spectral GF, Eq.~(\ref{sol-gfd}). The departures from  the $x \leftrightarrow 2-x$ symmetry are clearly visible 
in Fig.~(\ref{fig:rys3}), where we show the density of states for $x=0.2$, 0.5, and the symmetry related 
values $(2-x)=1.8$, 1.5. The comparison of the curves obtained for the pairs of various 
$x$ (0.2 and 1.8, and 0.5 and 1.5, respectively) shows that the differences are largest for energies close 
to the chemical potential $\mu$, {\it i.e.}, the point where the Kondo effect is expected. The differences at other energies 
seem to be related to those at $\mu$ by the sum rule, $\int_{-\infty}^{+\infty} dEN(E)=1$, which is always fulfilled with an 
accuracy better than $1\%$.

To obtain the above results, Fig.~(\ref{fig:rys3}), we have used the formulae (\ref{sol-gfd}) and (\ref{sol-id}),
which have been obtained, see the Suppl. Material \cite{sm}, using the decoupling scheme I. 
As discussed there, we have tried several different decouplings. The others,
{\it i.e.}, II and III, overall lead to the same behavior with 
small quantitative changes only, hence we are not showing the results for them here. For the discussion 
of decouplings II and III, and also a calculation scheme different from that presented in 
Sec.~\ref{sec:GFs}, see App.~\ref{app:other-dec} and App.~\ref{app:matrix} below. 

Interestingly, despite the asymmetry of the spectral GF, the average charge density per
spin is symmetric under $x\leftrightarrow 2-x$.
For a symmetrically coupled ($\Gamma^L=\Gamma^R$) quantum dot in equilibrium, the expression for the 
average occupation reduces to an integral of $N(E)$ weighted with the Fermi-Dirac distribution function.
The dependence of $\langle n_{\sigma}\rangle=\langle n\rangle$ on $x$ is shown  
in the inset to Fig.~(\ref{fig:rys3}) for $U=8$ and three values of $\varepsilon_d$, namely $-2$, $-4$, and $-8$.
We see that the symmetry is obeyed with an accuracy of $\approx 0.01$, which is
only slightly larger than the accuracy of the iterative computation: In the iterative process, 
$\langle n\rangle$ is used as a check of the accuracy of the solution, and we terminate the iteration when the change 
in $\langle n\rangle$ is less than 0.001 in consecutive steps. 
\begin{figure}
\includegraphics[width=0.9\linewidth]{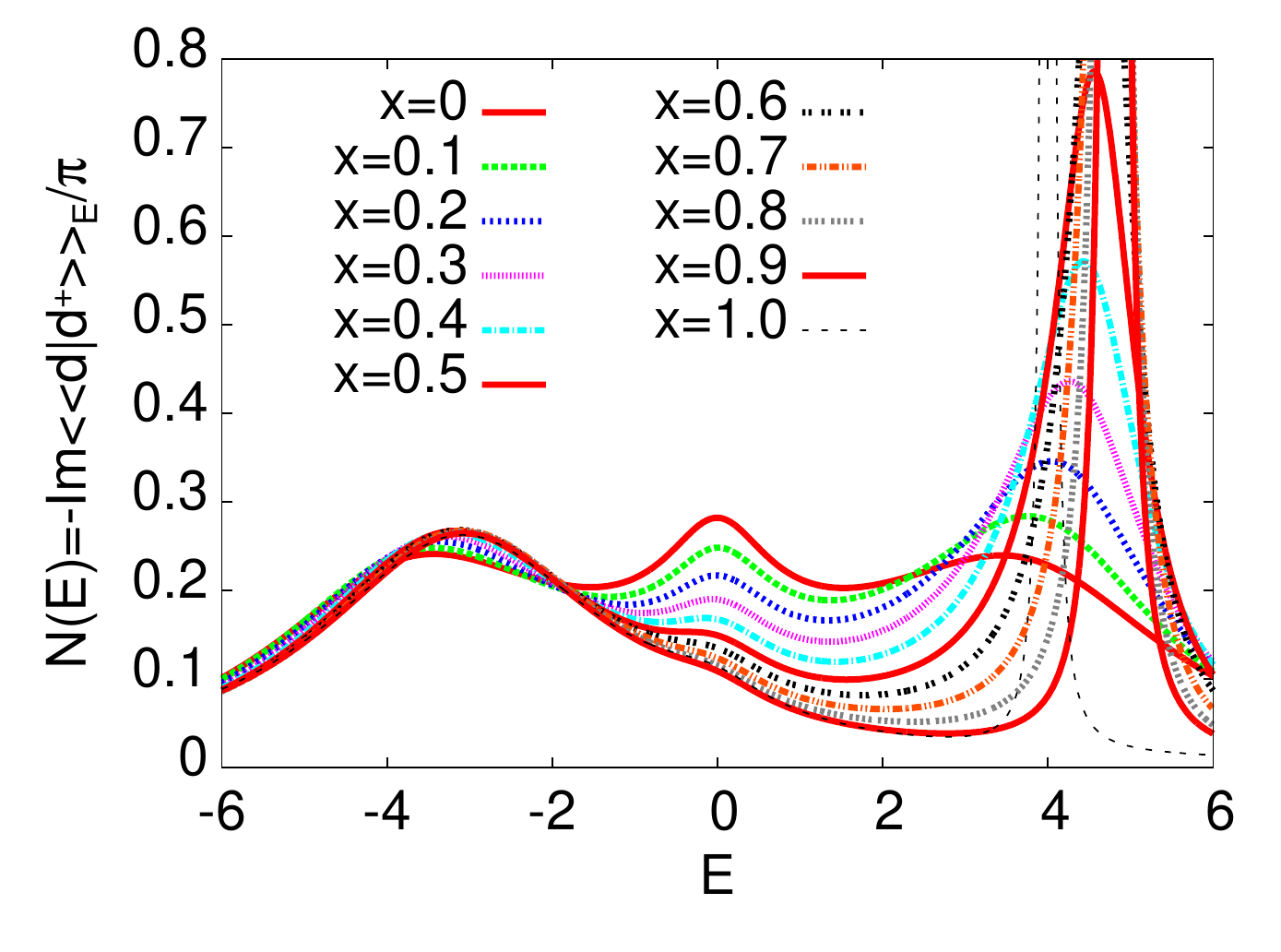}
\caption{(color online) The dot's density of states for a number of
$x$ values. Note the weak changes of the lower Hubbard band with $x$, and the 
opposite behavior for the upper one. The other parameters are $\varepsilon_d=-4$, $U=8$, and $T=0.3$. }
\label{fig:rys4} 
\end{figure}

Anticipating the discussion in the next subsection, we expect in fact that the
calculation of the spectral GF in the interval $0<x<1$ is, in the present approximation, more reliable 
than the results obtained for $1<x<2$. Hence we focus on the former regime, and illustrate in
Fig.~(\ref{fig:rys4}) in more detail the changes of the 
density of states with increasing $x$ for the particle-hole symmetric case, $\delta=0$. 
The general trends in the spectral GF for arbitrary $\delta$ are similar to those observed for the transport GF. 
The modifications of the lower Hubbard band are relatively small, while the Kondo peak and the upper
Hubbard band are strongly modified with increasing $x$. The Kondo resonance disappears, and 
the upper Hubbard band gets narrower and higher with its center shifting initially towards higher energies.

\subsection{Remarks on the asymmetry of the spectral GF}
\label{sec:asymmetry}
\begin{figure}
\includegraphics[width=0.85\linewidth]{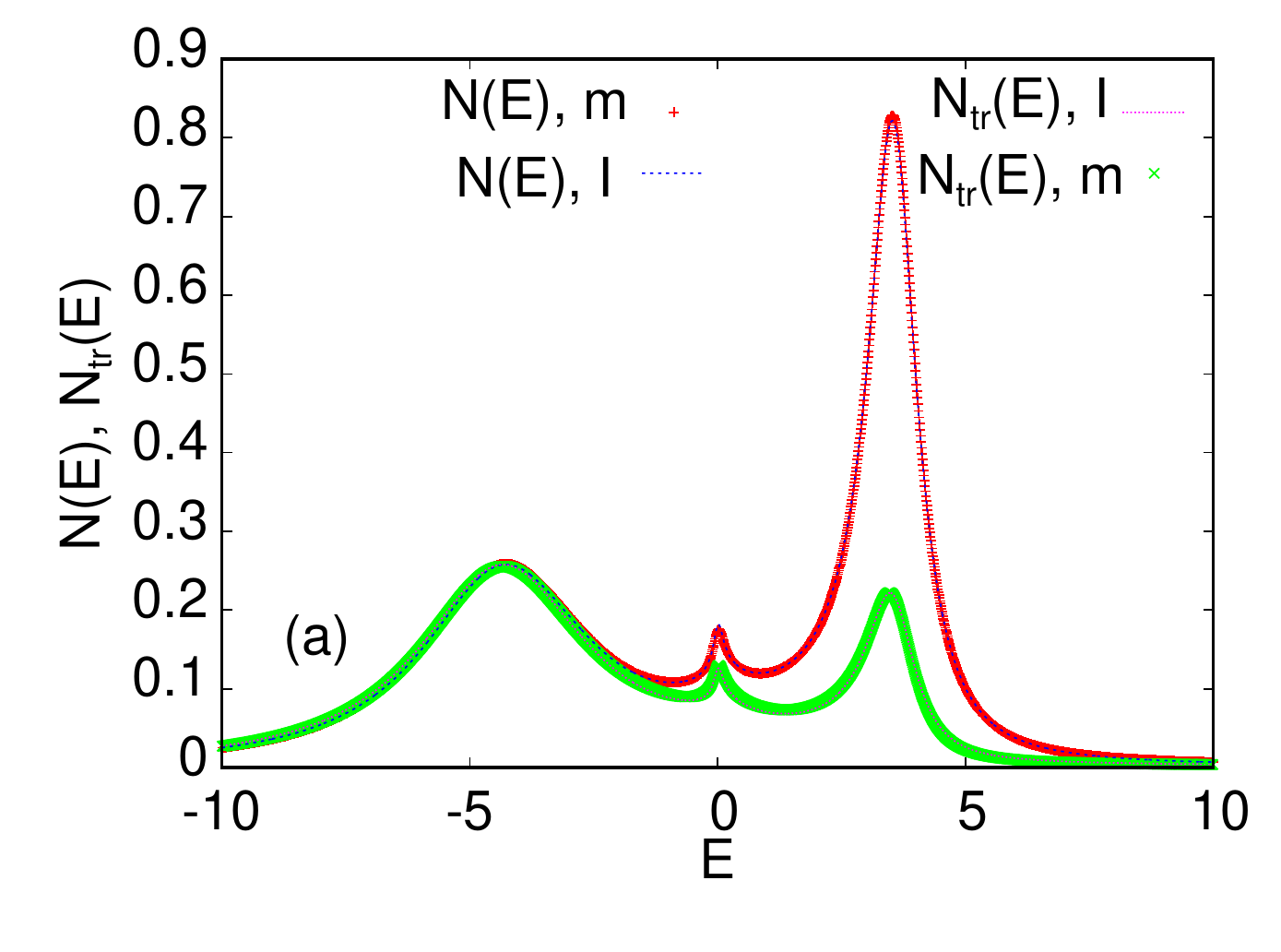}
\includegraphics[width=0.85\linewidth]{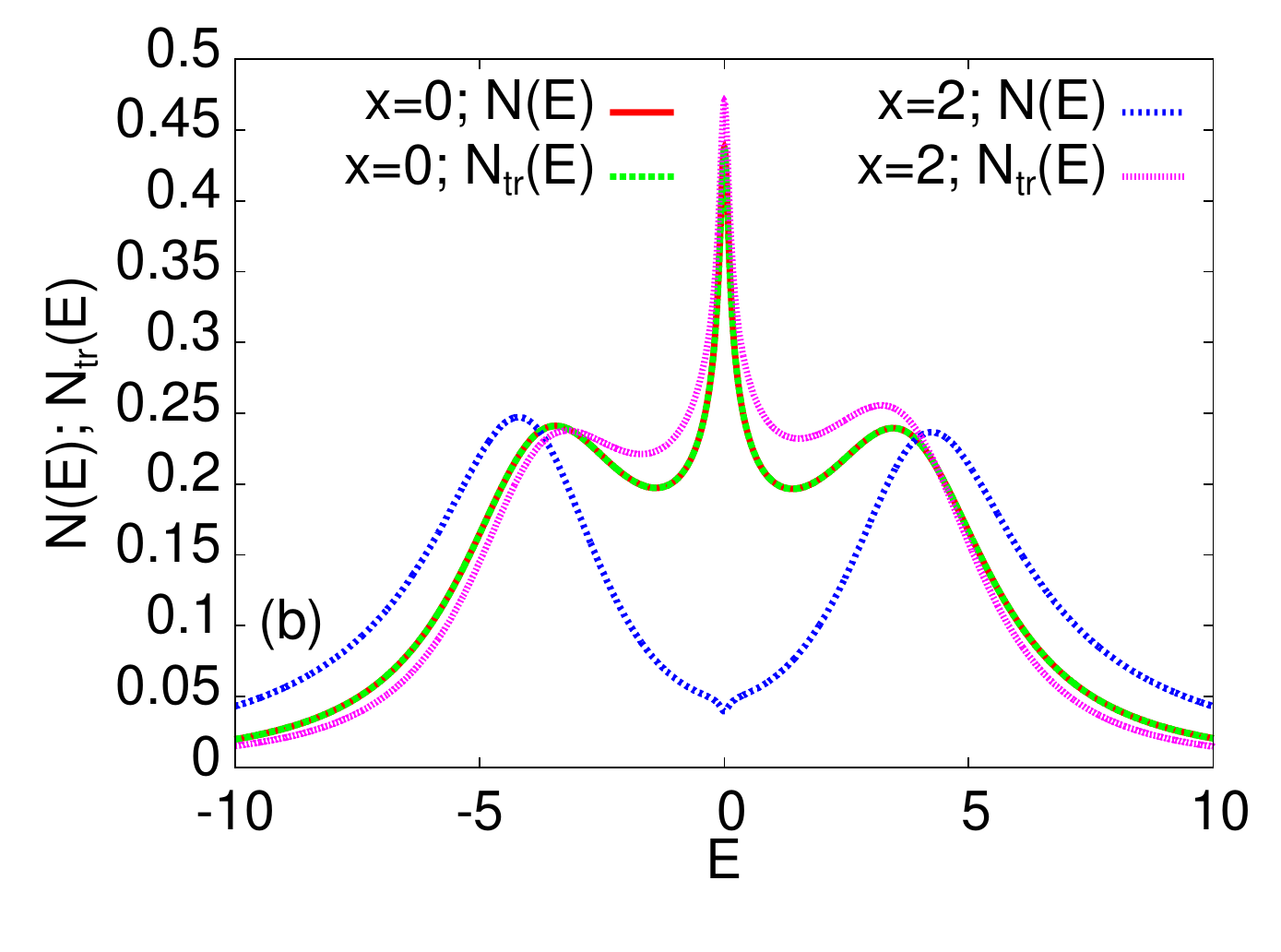}
\caption{(color online) (a) Comparison of the spectral and transport densities of states 
for $x=0.5$ and $\varepsilon_d=-5$, $U=8$, and $T=0.03$ calculated by two methods:
`m' stands for the matrix method, while `I' for decoupling I. Panel (b) illustrates the 
symmetry between these quantities for the particle-hole symmetric system 
with $\varepsilon_d=-4$, $U=8$, and $T=0.3$ calculated by the matrix method. For the full
discussion of this figure, see App.~\ref{app:matrix}.}
\label{fig:rys8} 
\end{figure}

A careful look at the spectral GF for $x=2$ in Fig.~(\ref{fig:rys8}) shows that a small dip 
appears at the Fermi energy ($\mu=0$). 
Such a dip in the energy dependence of the equilibrium density of states at the chemical potential 
is a characteristic feature of all previous decouplings \cite{kashcheyevs2006,sierra2017} for the standard Hubbard model, 
({\it i.e.}, for $x=0$). The only approach which cures the deficiency is that of Lavagna~\cite{lavagna2015} 
for the Hubbard model, which is applied here for the correlated-hopping model. Why does the approach fail
at $x=2$? To elucidate the reason why Lavagna's approach is not effective for the spectral function 
at $x=2$, we have to recall that the existence of the Kondo resonance 
for the Hubbard model in her approach is intimately related to the lifetime effects, 
{\it i.e.}, the use of the parameters $\tilde{\gamma}^{(1)}_{\sigma}$, $\tilde{\gamma}^{(2)}$ 
as discussed earlier~\cite{lavagna2015}.

First we note that the explicit dependence on $x$ in the formulae (\ref{a-sol-gf-t}) and (\ref{a-sol-sp2-t}) 
for the transport GF is through the factors $x(2-x)$ or $(x-1)^2$. The former vanishes for $x=0$ and $x=2$, 
and the latter is symmetric with respect to $x=1$. On the other hand,  
the inspection of the formula (\ref{sol-gfd}) for the spectral GF shows that for $x=0$ all
extra terms vanish, while for $x=2$ they do not but rather give a large contribution to
the self-energies $b_{1\bar{\sigma}}, \Sigma^T_{1\bar{\sigma}}$, and 
$b_{2\bar{\sigma}}, \Sigma^T_{2\bar{\sigma}}$. These self-energies at low $T$ lead to logarithmically divergent 
contributions close to the Fermi energy. For $x=0$, however, the divergent contributions from those terms which remain 
in $B_d$ and $n^d_{\mathrm{eff}}$ are cut off by the lifetime effects. On the other hand, this is not the case for $x=2$,
and a number of  diverging self-energies remain. 

With this insight, we expect that in order to obtain the correct symmetry of the spectral GF one has
to calculate, instead of projecting, those GFs which contain two lead operators.
A careful inspection of the decoupling procedures shows, in fact, that the problem lies in the vanishing of certain contributions for $x=0$ but not for $x=2$. Thus, in order to render the expression for 
$\langle\langle d_{\sigma}|d^{\dagger}_{\sigma}\rangle\rangle_{\omega}$ symmetric, extra terms are needed.
These can only result from higher-order contributions to the $S_{n,d,c}^{\mathrm{sp}}$ terms.  As already noted, 
the calculations of these ``missing'' terms can, in principle, be performed in full analogy 
to previous calculations for the Hubbard model~\cite{vanroermund2010}, but for the model at hand 
they are very complicated and will not be pursued here. Thus we conclude that contrary to the Hubbard
model, where lifetime effects~\cite{lavagna2015} can mimic the role of the fourth order 
terms~\cite{vanroermund2010}, the symmetry of the spectral function of the correlated-hopping model 
requires calculations up to fourth order in the tunneling amplitude.

\section{Transport characteristics}
\label{sec:transport}
\subsection{Charge and heat conductance, and thermopower: linear regime}
\begin{figure}
\includegraphics[width=0.9\linewidth]{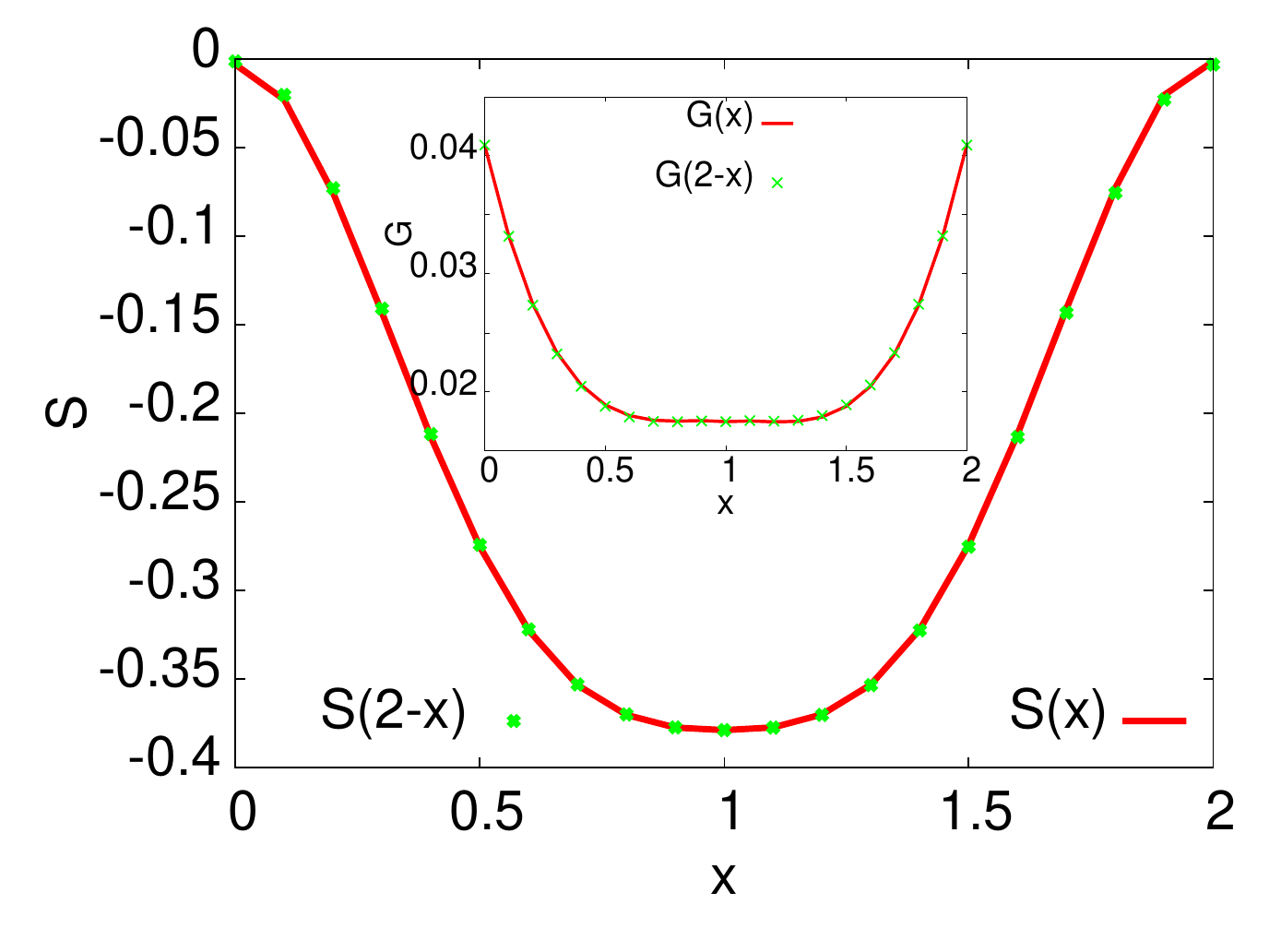}
\caption{(color online) The linear thermopower $S$ (main panel)
 calculated for $\varepsilon_d=-4$, $U=8$, and $T=0.3$
{\it vs.} $x$. The inset shows the $x$-dependence of the linear conductance $G$.
The solid lines indicate the thermopower (conductance) calculated for $0<x<1$, while the points 
correspond to the values obtained for $0<2-x<1$. The symmetry is nearly perfect in both cases. }
\label{fig:rys5} 
\end{figure}

\begin{figure}
\includegraphics[width=0.85\linewidth]{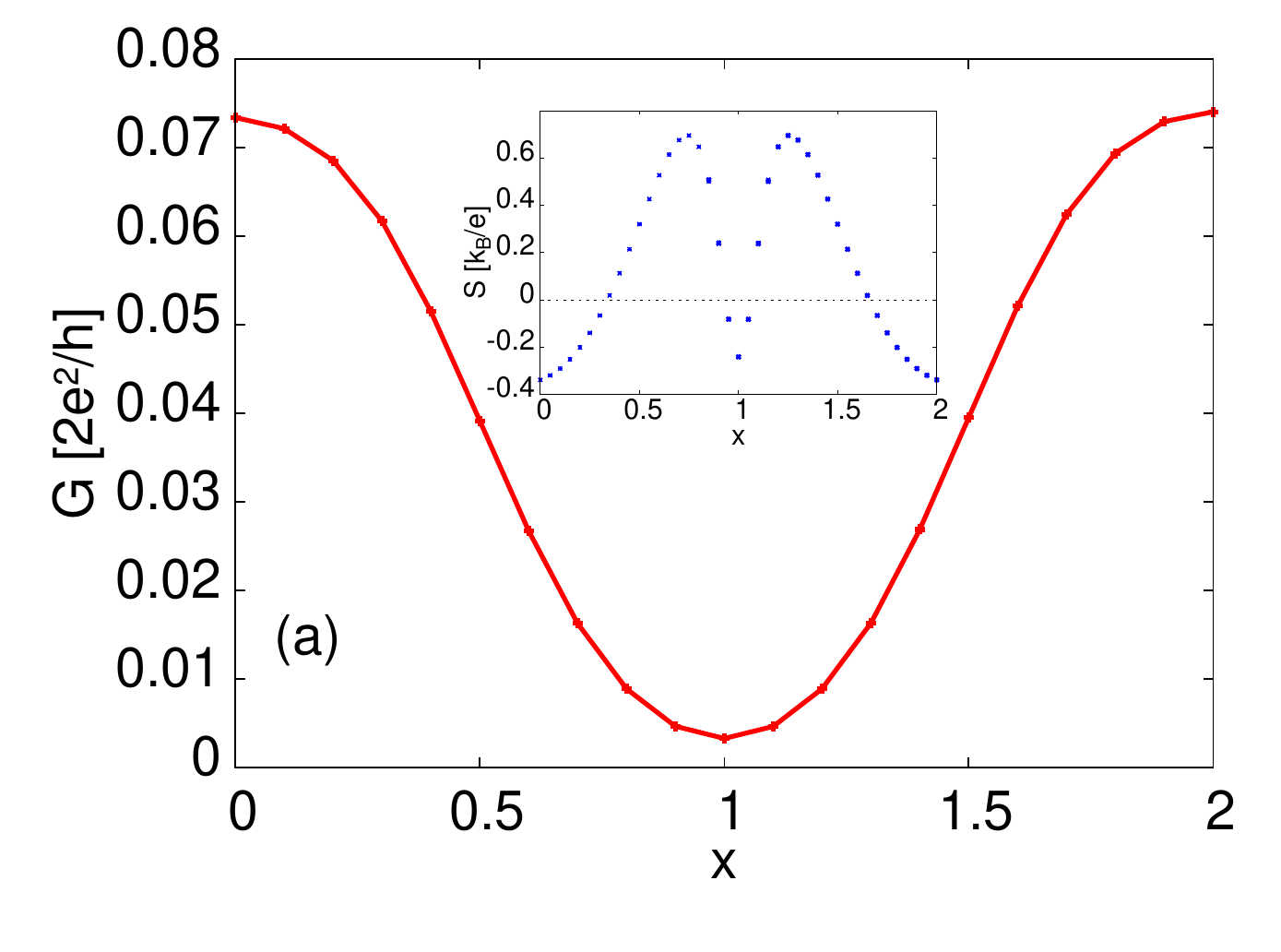}
\includegraphics[width=0.85\linewidth]{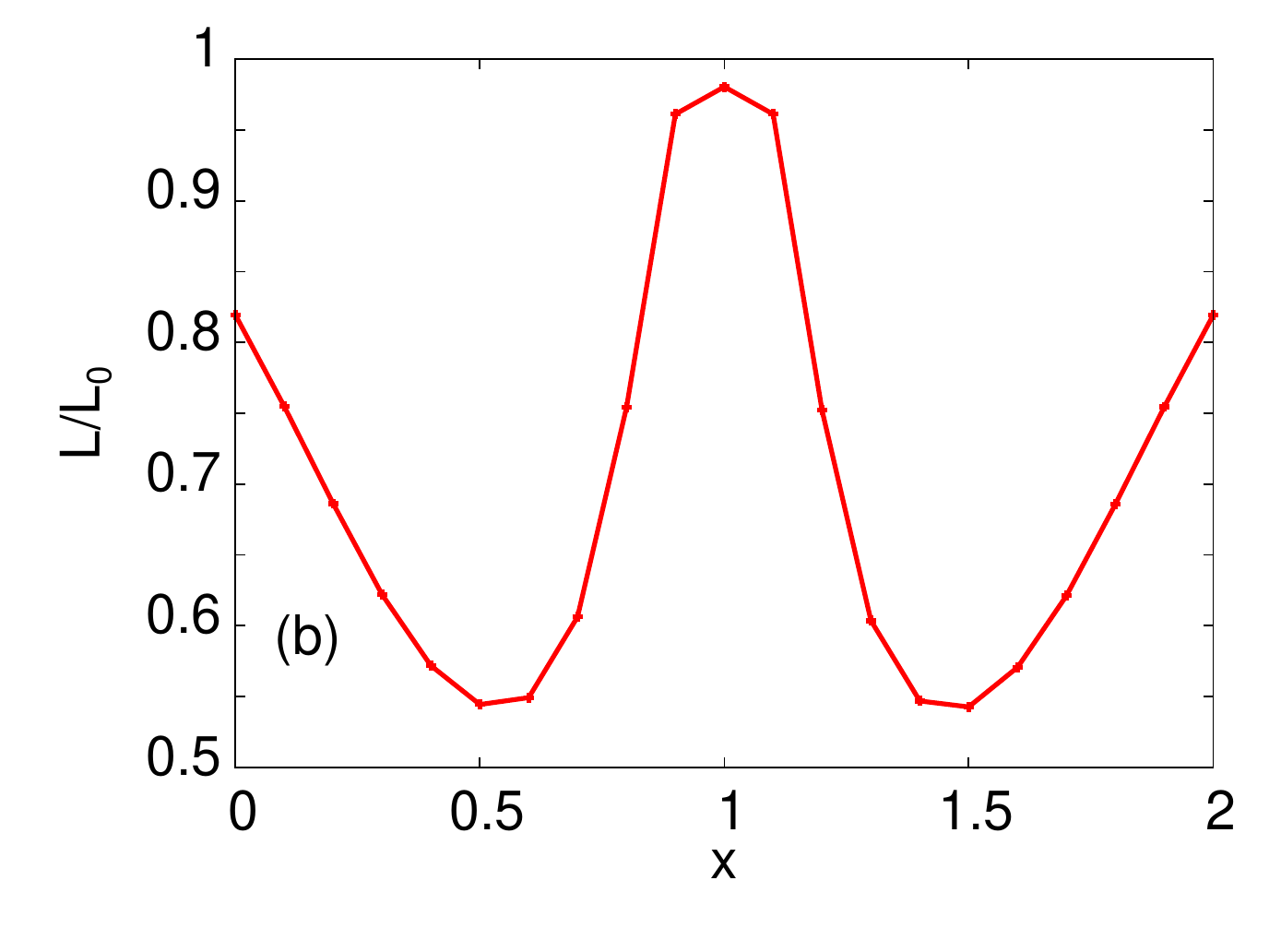}
\caption{(color online)  
Conductance and thermopower $S$ (inset), as well as $L/L_0$ {\it vs.} $x$ as calculated 
for $\varepsilon_d=-8$, $U=8$, and $T=0.3$.}
\label{fig:rys6} 
\end{figure}

We emphasize again that all transport coefficients depend on the transport GF only,
and hence fulfil the required $x$ symmetry. We focus the following discussion on the conductance $G$,
the thermopower $S$, and the Wiedemann-Franz ratio $L=\kappa/(GT)$. In agreement with the preceding 
discussion, the nearly perfect $x$ {\it vs.} $(2-x)$ symmetry is very well visible in $G$ and $S$.
In Fig.~(\ref{fig:rys5}) we show the linear conductance $G$ and the thermopower 
$S$: the comparison of these quantities {\it vs.}
$x$ (solid curves) and {\it vs.} $(2-x)$ shows that the symmetry is well
obeyed. In Fig.~(\ref{fig:rys5}) the symmetry is illustrated 
for the model with $\delta=0$, but it is also valid for arbitrary values of this parameter. 
For non-vanishing values of $\delta$ the functional dependence of $G(x)$ {\it vs.} $x$ differs 
but the symmetry remains intact. Interestingly, the overall dependence 
of the conductance on $x$ shown in Fig.~(\ref{fig:rys6}) is similar to that in the previous figure. 
In both cases the conductance takes on 
maximal values for $x=0$ and $x=2$. However, it has to be noted that the behavior of $G(x)$ 
in the particle-hole symmetric case (shown in Fig.~(\ref{fig:rys5})) is due to the  
destructive effect of $x$ on the  Kondo peak at the chemical potential. 
An increase of $x$ destroys the Kondo resonance, which leads to a smaller conductance. 
On the contrary, for the parameters of Fig.~(\ref{fig:rys6}) it is the upper band of the
transport density of states which is located close to the chemical potential that 
gives the largest contribution to $G$. The modification of this
part of the transport spectrum (as visible in Fig.~(\ref{fig:rys6a}a)) determines the $x$ dependence of $G$.

The thermopower dependence on $x$ in the above two models is more complicated. In the first 
case, $S$ attains zero values for the perfectly
symmetric transport densities of states at $x=0$ and $x=2$, and decreases for $x$ tending towards 1. 
The complicated sign changes of $S$ for the model with $\varepsilon_d=-8$, 
shown in Fig.~(\ref{fig:rys6}a), can be approximately read off from the slope of the transport  
density of states shown in Fig.~(\ref{fig:rys6a}a). 
This is due to the fact that in the linear regime $S$ is proportional to the derivative of $N_{\mathrm{tr}}(E)$ taken 
at the chemical potential; {\it cf.} Eq.~(\ref{mott-s}). We have also calculated the thermal conductance, 
and its dependence on $x$ traces the charge conductance. Thus we are not showing the plots here. Instead
in Fig.~(\ref{fig:rys6}b) the Wiedemann-Franz ratio $L=\kappa/(TG)$ normalised to the Lorenz number 
$L_0=(\pi^2/3)(k_B/e)^2$ is presented ($\kappa$ denotes the thermal conductance). One observes that for 
the parameters used the ratio is smaller than unity. This indicates a non-Fermi-liquid 
behavior of the system with hampered charge transport.

One of the most interesting questions related to the present study is the identification of
measurable consequences of correlated hopping. In this context, it should be emphasized that one has experimental
control over virtually all parameters of the devices under discussion. In particular, the gate bias
independence of the parameters $\Gamma^\lambda$ demonstrated recently \cite{dutta2019} for QDs
fabricated using the electromigration technique, supports the hope to achieve this goal.

Naturally one would like to measure some transport characteristics of the system and infer the information 
about the actual value of $x$. The  symmetry with respect to changing $x$ cannot serve the purpose, 
as this parameter most likely is beyond experimental control. However, other characteristics of the
transport coefficients come to mind: namely the existence of the distinctive peaks in the conductance,
and the concomitant saw-tooth shape of the thermopower.
\begin{figure}
\includegraphics[width=0.85\linewidth]{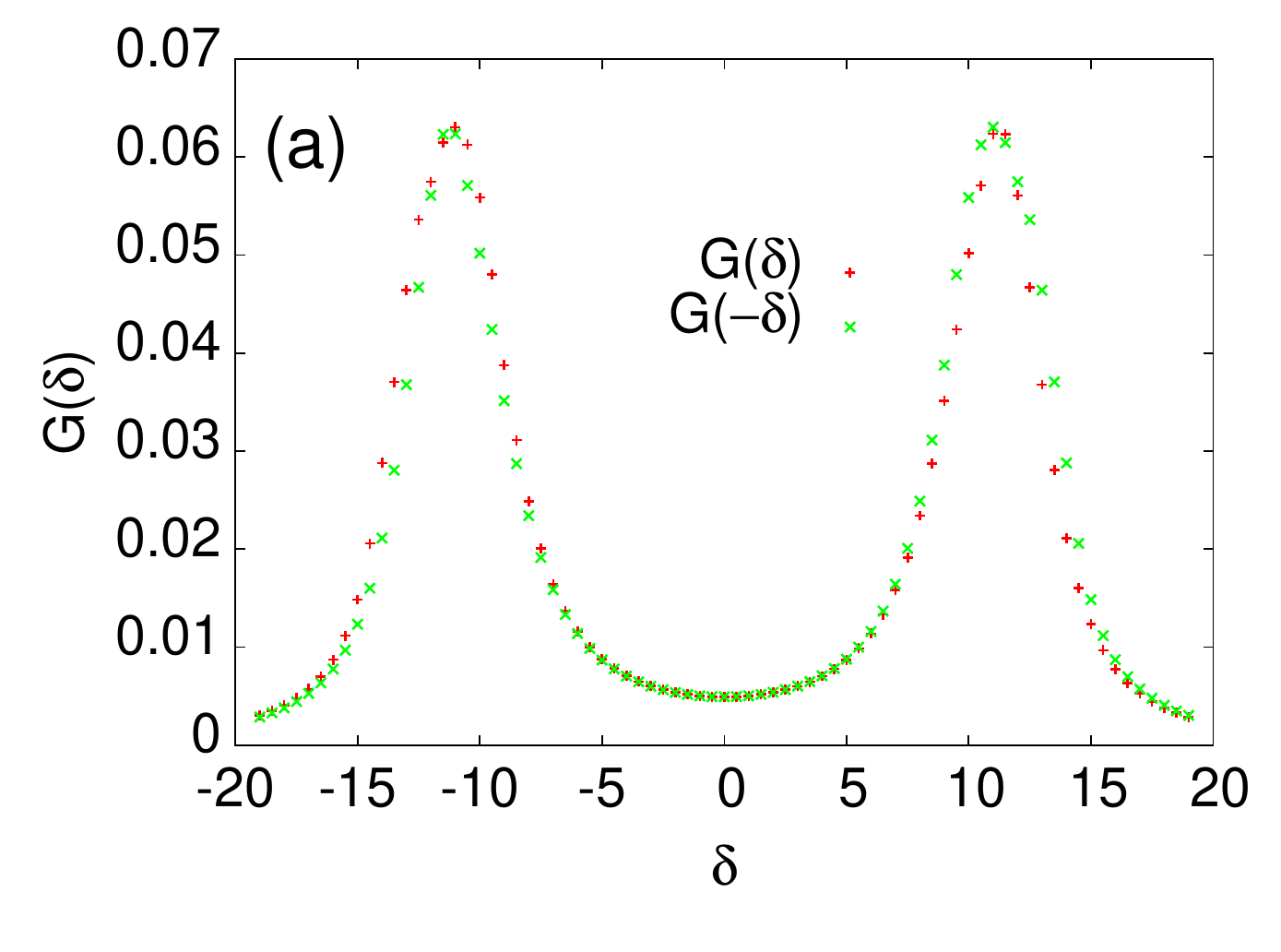}
\includegraphics[width=0.85\linewidth]{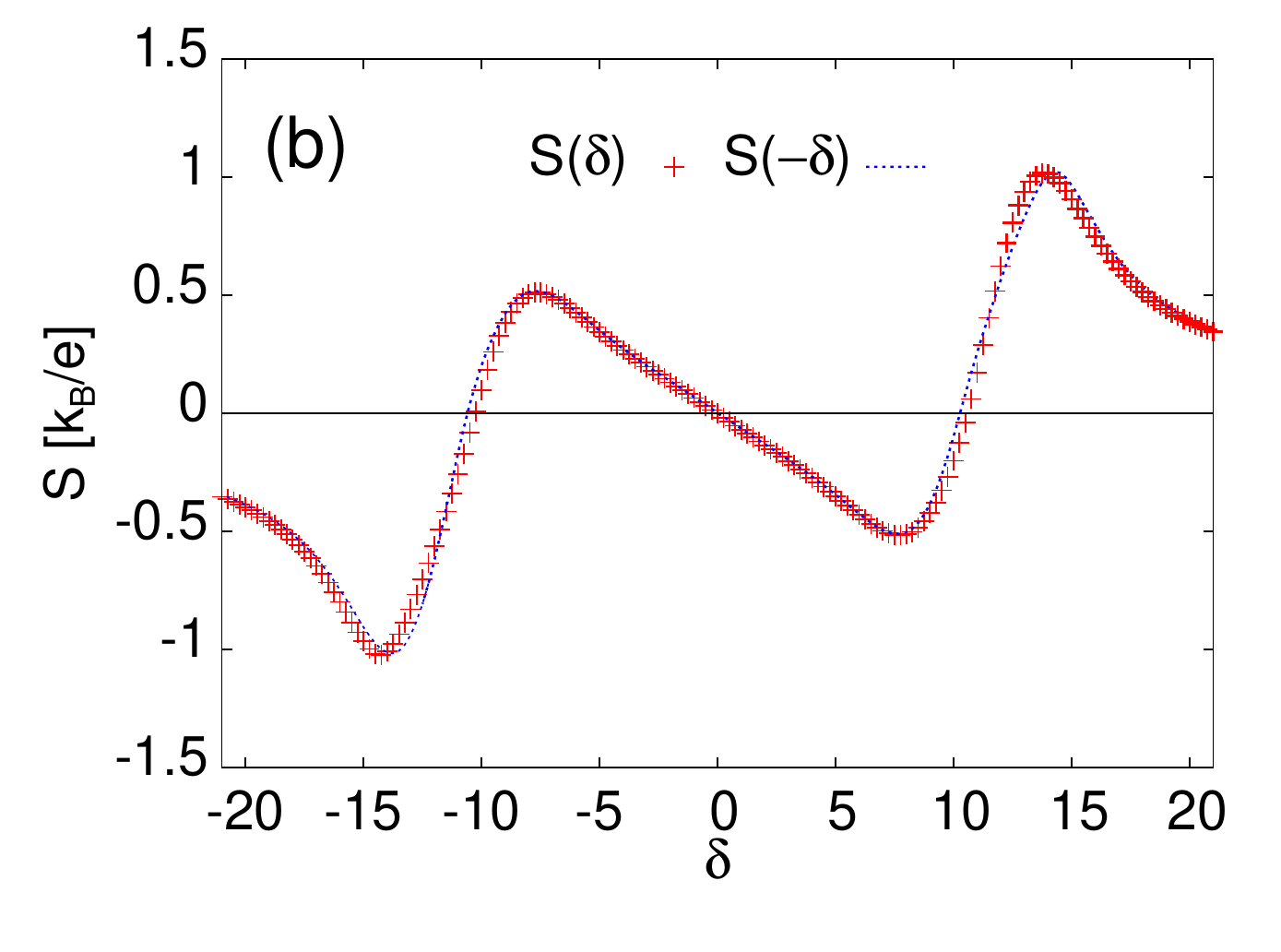}
\caption{(color online) Symmetry of the transport coefficients with respect to gate bias 
(characterized by $\delta =\varepsilon_d+U/2$) for the model with $x=0$. 
Panel (a) shows the conductance, and (b) the thermopower. Other parameters
are $U=22$, $T=0.5$.}
\label{fig:rys11} 
\end{figure}
The answer depends on whether one considers the linear or the strongly non-linear transport regime. 
In the linear case (we are interested in here), and for temperatures low on the scale of $\Gamma_0$, 
the transport parameters probe, as visible from Eqs.~(\ref{moments}), the region 
of energy of the width of a few $k_BT$ in the vicinity of the chemical potential (here $\mu=0$), so the
observed changes in the transport density of states are directly measurable (\ref{przew-lin-approx}).
We claim that it is the $x$ dependent relative height of the two conductance peaks of the single-level 
quantum dot which gives direct information on the correlated-hopping term. 
The observed maxima of $G$, measured as function of gate bias, are related to the corresponding 
maxima in the transport density of states: one peak builds up when the on-dot energy band around
$\varepsilon_d$ crosses the Fermi level ($\mu=0$), and the other when the upper Hubbard band centered 
around $2\varepsilon_d+U$ sweeps through $\mu$. For $x=0$ the two peaks are identical as visible in 
Fig.~(\ref{fig:rys11}a). Similarly the corresponding ``anti-symmetry'' is observed for the thermopower, 
as visible in panel (b) of Fig.~(\ref{fig:rys11}).  
This argumentation is applicable for a symmetrically coupled quantum dot, {\it i.e.}, for $\Gamma^L=\Gamma^R$.  

\begin{figure}
\includegraphics[width=0.85\linewidth]{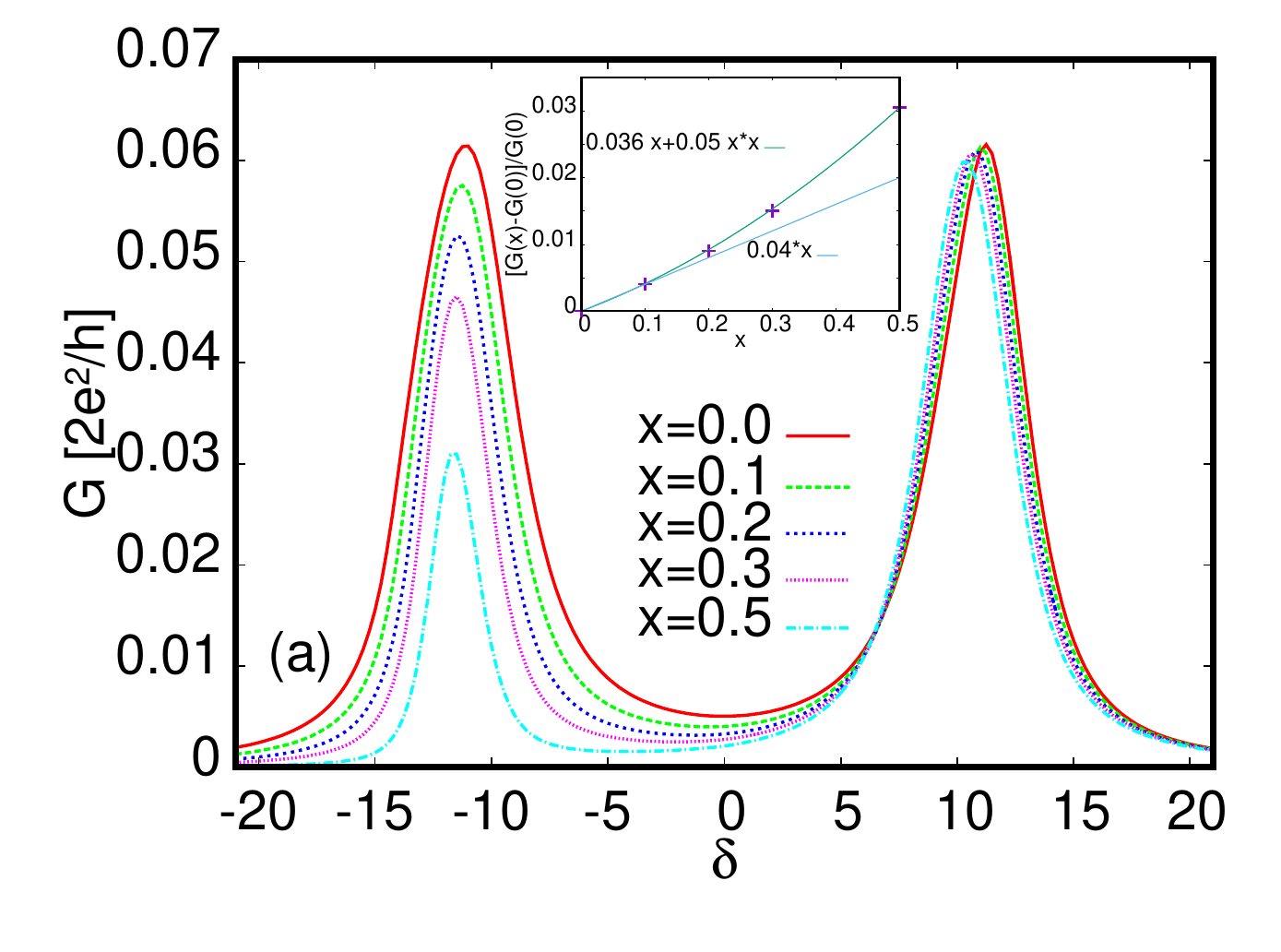}
\includegraphics[width=0.85\linewidth]{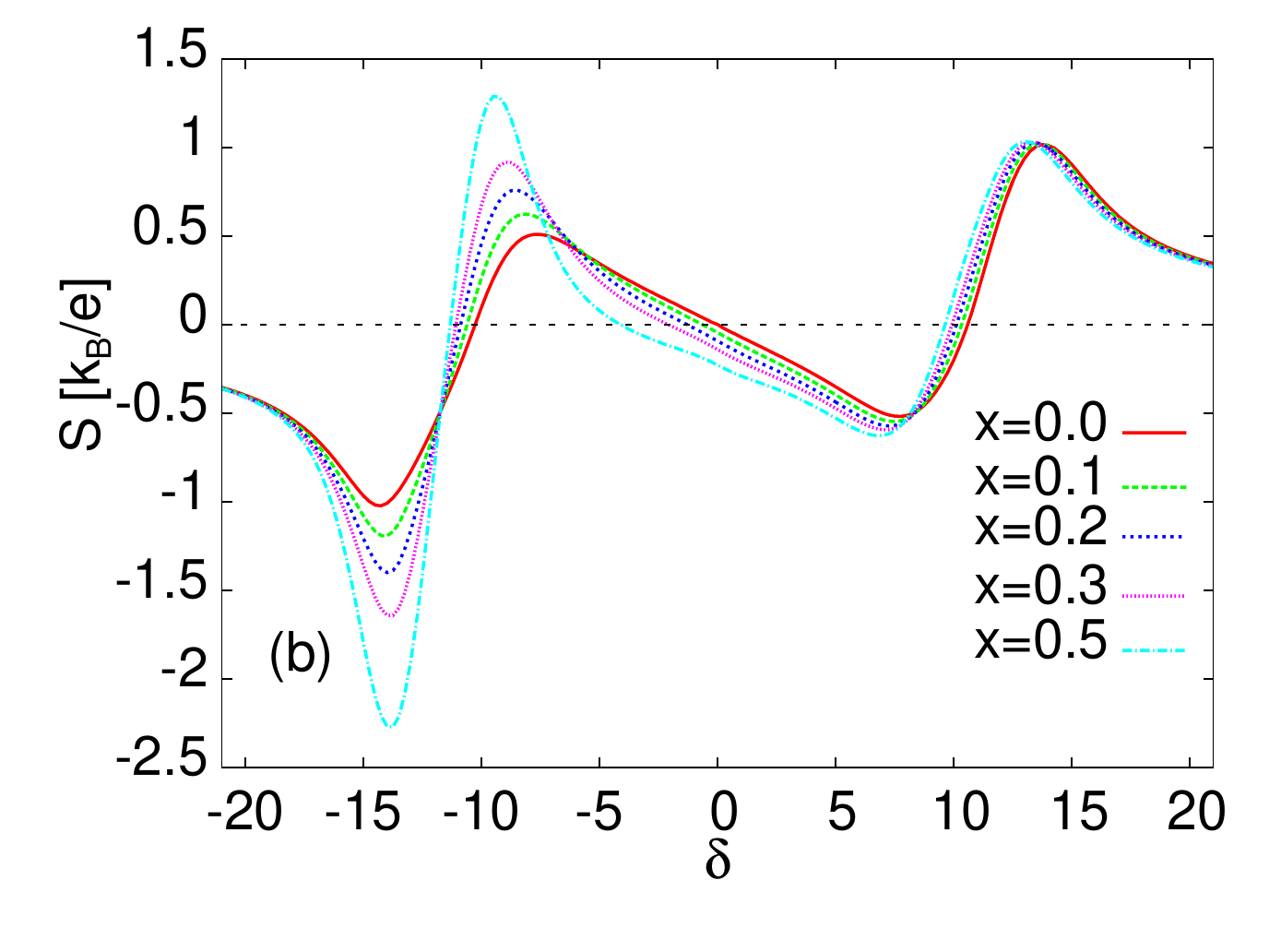}
\caption{(color online) The effect of $x$ variation on the gate bias dependence 
(characterized by $\delta =\varepsilon_d+U/2$) of the conductivity, (a),
and the thermopower, (b), for the model with $U=22$ at $T=0.5$. Increasing $x$ increases the asymmetry of the conductivity.
The upper peak remains essentially intact while the height of the lower one decreases. The inset shows that the decrease of the height is faster than linear.}
\label{fig:rys10_} 
\end{figure}

In Fig.~(\ref{fig:rys10_}a) we show the dependence of the conductance on gate bias, {\it i.e.}, on $\varepsilon_d$, 
for a few values of $x$, namely $x=0$, 0.1, 0.2, 0.3, and 0.5. One observes a change of the relative height
between the lower and the upper conductance ``bands''  with increasing $x$. The lower 
conductance peak decreases with $x$ while the upper one stays constant. The observed decrease 
is  faster than linear as shown in the inset to the figure. The proportionality factor in the linear fit,
here equal to 0.04, is not universal, but depends on the details. However, the decrease of the lower peak height with $x$ 
is a universal effect for a symmetrically coupled quantum dot, and gives immediate information on the 
very presence of the correlated-hopping contribution.

We emphasize that the observation of the different heights of the two consecutive conductance peaks
in the two-terminal quantum dot provides a unique proof of the existence of correlated hopping.
As we are studying a single-level quantum dot, the main condition related to experimentally studied devices 
is that the distance between consecutive 
levels in the dot has to be larger than the Coulomb repulsion $U$. Otherwise, the consecutive conductance peaks
would correspond to singly occupied levels. This probably is the most serious 
condition to fulfil. Additional information can be drawn from the analysis of 
the thermopower. However, the gate bias dependence of $S$ is slightly more complicated, as is visible from
Fig.~(\ref{fig:rys10_}b): An increase of $x$ leads to an increase of the amplitude of $S$
for $\delta$ values corresponding to the lower conductance peak, $\delta \approx -12$, while $S$ 
remains virtually unchanged for $\delta \approx +10$, corresponding to the upper conductance peak.

\begin{figure}
\includegraphics[width=0.85\linewidth]{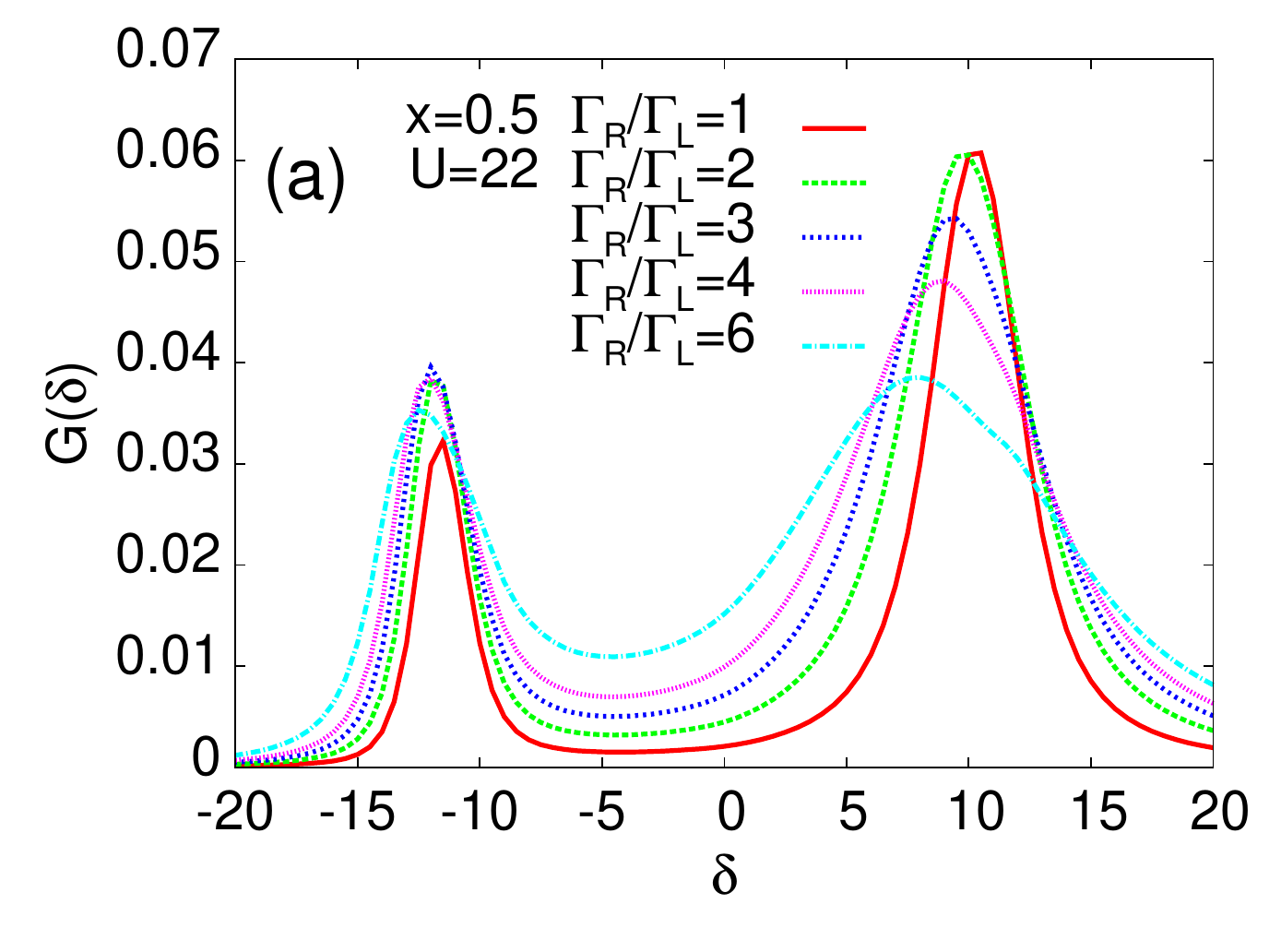}
\includegraphics[width=0.85\linewidth]{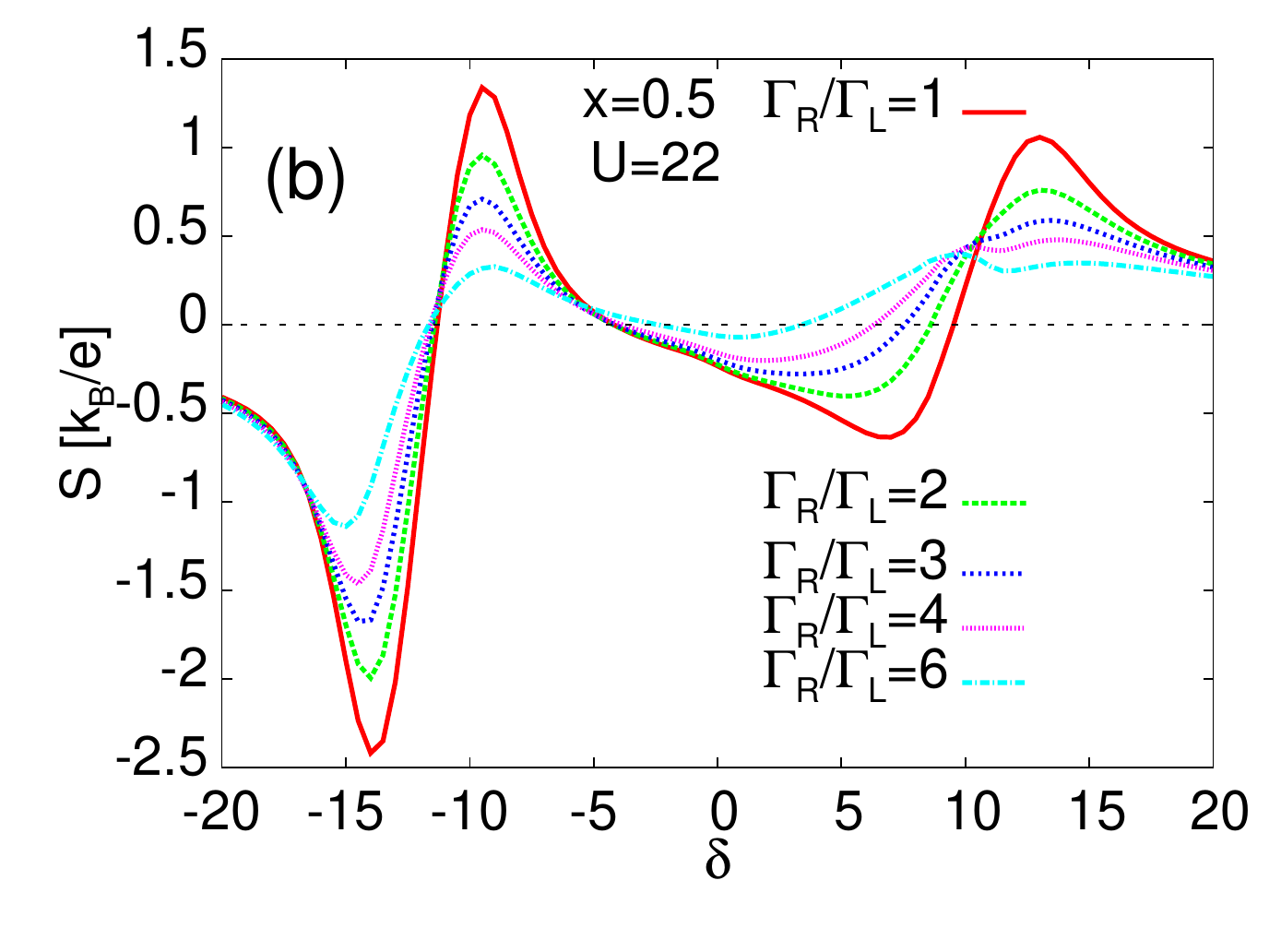}
\caption{(color online) Dependence of (a) the linear conductance and (b) the thermopower on gate bias, 
characterized by $\delta =\varepsilon_d+U/2$, for $x=0.5$, $U=22$, and $T=0.5$. 
One observes only a small effect of the anisotropy of the couplings on the lower conductance peak, but large 
changes of the upper peak, namely a decrease of its height and an increase of its width when increasing the
ratio $\Gamma^R/\Gamma^L$.} 
\label{fig:rys7} 
\end{figure}

The effect of non-symmetric couplings on the conductance and the thermopower is shown 
in Fig.~(\ref{fig:rys7}) for $\Gamma^R/\Gamma^L=2,3,4,6$. The anisotropy only weakly 
affects the lower conductance  peak. Its effect on the upper peak is appreciable and includes a decrease
of the magnitude and an increase of the width. The former effect masks the asymmetry in the peak
heights induced by finite $x$, hence its unique identification becomes more difficult. However, the 
thermopower (Fig.~(\ref{fig:rys7}b)) reacts in a more complicated way. 
Its overall amplitude diminishes in comparison to the symmetrically coupled dot, but both low and high $\delta$
parts are modified. The decrease of the magnitude of the thermopower variations can be understood by noting that $S$ is
proportional to the slope of the conductivity at the corresponding energy, which is known as Mott relation.

Our EOM results quantitatively agree with those obtained by the NRG technique \cite{lin2007,tooski2014}.
In particular, an increase of $x$ induces similar modifications of the conductance and thermopower in both
methods. This is well seen by comparing, {\it e.g.}, our Fig.~(\ref{fig:rys10_}) with panels (a) and (b) in 
Fig.~3 of \cite{tooski2014}. It is more difficult to directly compare our results with those 
presented in \cite{lin2007}, as these authors concentrate on such aspects as spin conductance and
temperature dependencies. However, some curves shown in their Fig.~3 are close to our results
for the corresponding set of parameters.

As a brief intermediate r{\'e}sum{\'e}, we note that
all measurable characteristics exhibit the required symmetry properties. In particular, we have argued  
that a detailed experimental analysis of the gate bias dependence of both conductance and thermopower,
in devices without orbital degeneracy and such that an adequate control of the symmetry of the couplings is
feasible, may elucidate the role played by the correlated hopping, and may even allow for the extraction
of $x$.

\subsection{Non-linear conductance}\label{sec:nl}
The non-equilibrium Green function approach is well suited to treat finite voltages, since the EOM captures, 
albeit in a not well controlled way, higher-order scattering processes including those analysed in \cite{gergs2015}.
However, the full analysis of the conductance and other transport characteristics in the non-linear regime is
beyond the scope of the present paper: these quantities not only depend on $x$, $\delta$, and $V$, but also 
on temperature, Coulomb interaction, and the anisotropy of the couplings.

\begin{figure}
\includegraphics[width=0.85\linewidth]{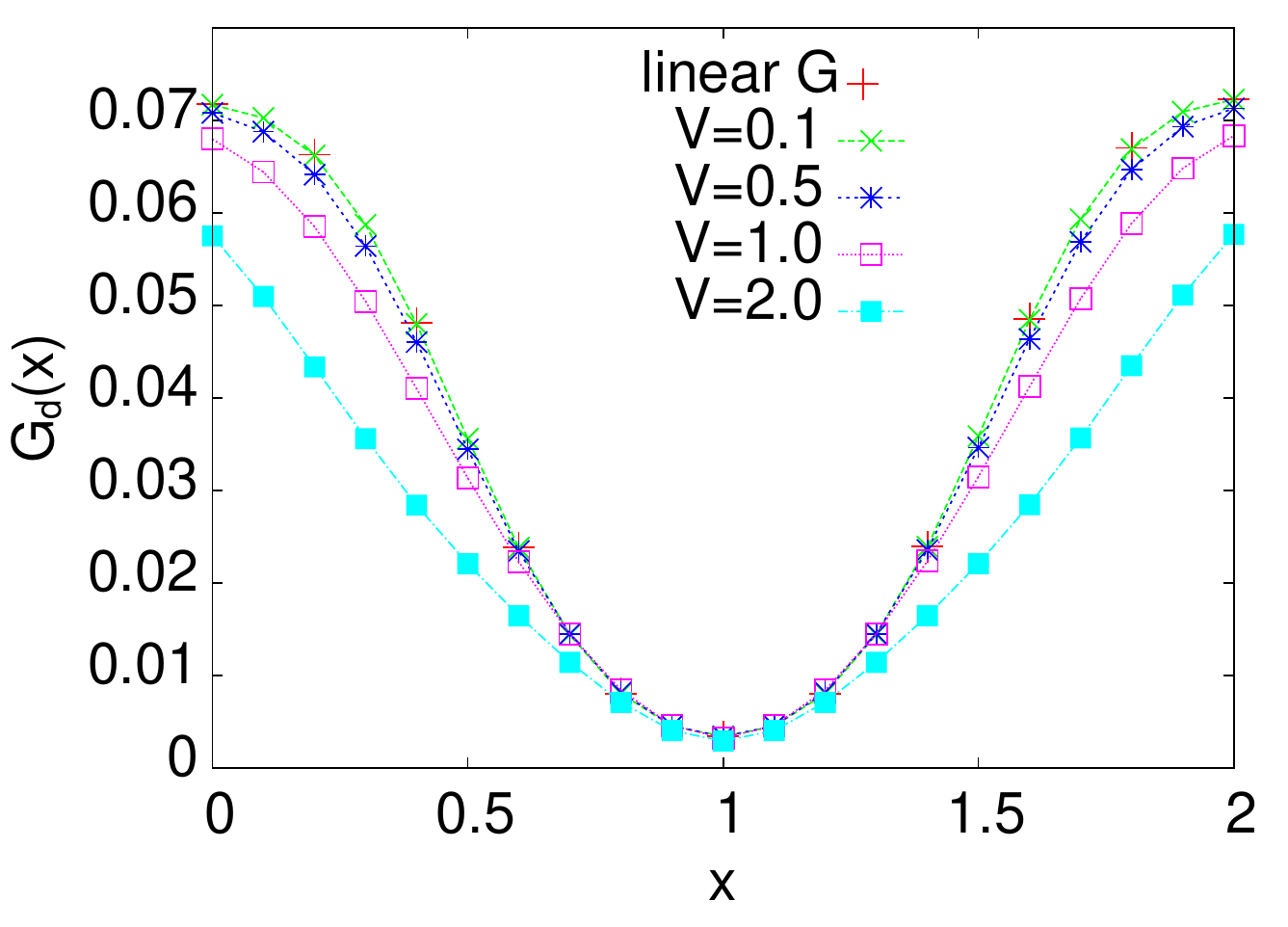}
\includegraphics[width=0.85\linewidth]{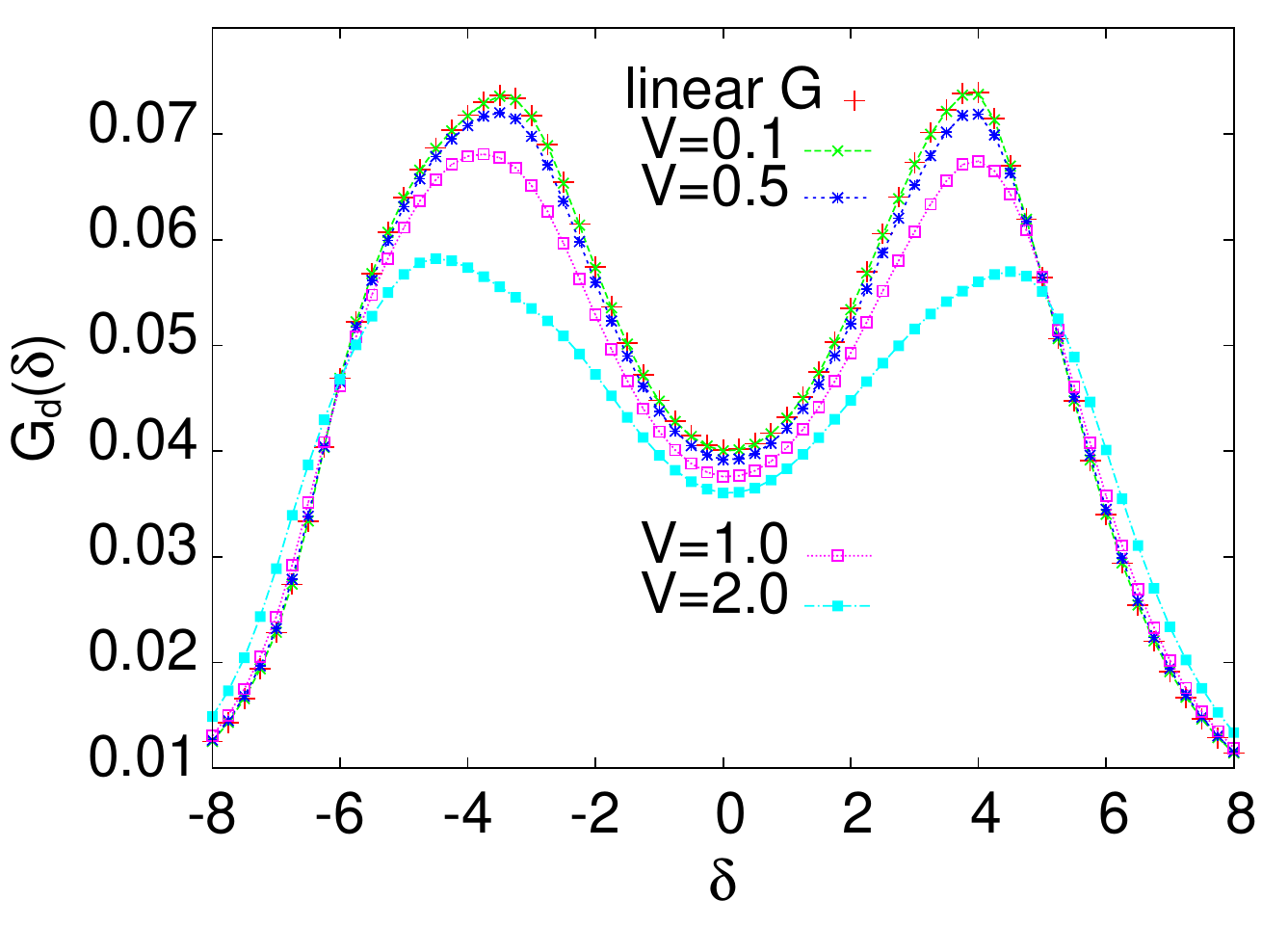}
\caption{(color online) (a) Differential conductance $G_d$ {\it versus} $x$ for $\delta=-4$, $U=8$, $T=0.3$, and a few values of the voltage $V$ as indicated. Panel (b) shows the dependence of $G_d$ on $\delta=\varepsilon_d+U/2$ for $x=0.3$ and the same $U$ and $T$. In both panels the linear conductance is shown by red pluses.} 
\label{fig:rys12} 
\end{figure}

Here we focus on the
differential conductance, $G_d(x,\delta, V)= {\partial I}/{\partial V}$. We present results for
the dependence of $G_d$ on $x$ for $\delta=-4$ and a number of voltages $V$ (Fig.~(\ref{fig:rys12}), panel (a)), and
the dependence of $G_d$ on $\delta$ for $x=0.3$ (panel (b)). In both cases $U=8$ and $T=0.3$. The linear conductance, 
formally corresponding to $V=0$, is shown by red pluses. For a small voltage, $V=0.1$, the differences between
small-voltage and zero-voltage results are small, but
they strongly increase with $V$. The departures from the linear regime are more pronounced for small $x$ 
(and $2-x$), and close to the resonant values of the gate bias when the conductance is maximal. 

For small values of $x$ and $2-x$, the changes of $G_d$ with $V$ are relatively large, but decrease for $x$ 
approaching $x=1$, see panel (a) in Fig.~(\ref{fig:rys12}).
These variations can be understood by recalling the full formula (\ref{eq:c-curr-wbl})
for $I$, and noting that the spectral Green function entering it also depends on the voltage $V$ in a rather complicated 
way. For a finite voltage, the energy integration interval depends on temperature, but generally is of order $V$ at 
low $T$, thus the actual value of the current and the differential conductance depend on the detailed behavior of the 
transport spectral density in the considered energy interval. For $x=1$, the differences are very small, due
to the rather smooth dependence of the transport spectral density on energy in the region between $\mu_L$ and 
$\mu_R$ ({\it cf.} Fig.~(\ref{fig:rys6a})). It is worth noting that the $x \leftrightarrow 2-x$ symmetry of 
the conductance is valid also in the non-linear regime.

The asymmetry between the conductance peaks for positive and negative values of $\delta$, visible in panel (b) of
Fig.~(\ref{fig:rys12}), are related to the particle-hole symmetry breaking by the correlated hopping. The 
asymmetry depends on the actual value of $x$ and may thus be utilized in precise experiments 
to obtain information on its value.

\section{Summary and conclusion}
\label{sec:summconcl}
The transport of charge and heat {\it via} nanostructures consisting of a quantum dot (QD) coupled to two
external normal electrodes has been studied by the non-equilibrium Green function (GF) technique in combination
with the equation of motion (EOM) approach. The system is described by the Anderson Hamiltonian containing 
not only the standard single-particle tunneling term, but also an additional one 
of many-body origin, known in the literature as assisted or correlated hopping. 
This term, often neglected in the analysis of transport measurements of QD nanostructures,
breaks the particle-hole symmetry of the model and thus may
be important or even decisive for the interpretation of various experiments. 

The  correlated-hopping term, which we have characterized by the
parameter $x$ ($0 \le x \le 2$), modifies the tunneling part of the single-impurity model.
Employing the non-equilibrium GF method to describe charge and heat transport, it becomes apparent 
that two different Green functions 
are  needed: one of them, which we denote {\it transport} GF, enters the formulae for the charge and heat flux, 
while the other is related to the dot density of states and hence the dot's average occupation.
However, the equations for both GFs are coupled to each other {\it via} the 
occupation and certain self-energies. To obtain each of the GFs within the EOM technique, one has to
project higher-order GFs, arising in the chain of EOMs, onto lower order ones. 
The simplest decoupling leads to a transport GF fulfilling the $x$ symmetry of the model.
However, the spectral GF is found {\it not} to be symmetric under $x \leftrightarrow 2-x$. 
Attempting to cure this deficiency, we have tried three different decoupling schemes 
(see \cite{sm}), but failed to achieve the required symmetry. Even the 
calculation of the full matrix GF with various decouplings did not restore the $x$-symmetry.
We argue that the symmetry restoration in the spectral GF $g^r(E)$ requires the inclusion of higher 
order GFs, thereby introducing higher powers of $x$. In contrast to the spectral GF, the
dot occupation and all transport coefficients preserve the proper symmetry. Hence we are confident that
the transport properties calculated in this work, and their parameter dependencies, are reliable. 

The analytical approach employed here clarifies explicitly that for the description of the model
with correlated hopping two characteristic Green functions are needed. One of them defines the transport 
through the system, and the other the thermodynamic properties, like the on-dot density of states or
the occupancy of the dot. The intimate coupling between both GFs is realized $via$ the on-dot occupancy 
of the opposite-spin electrons and various self-energies. This, in conjunction with the general formulae 
for the currents, clearly shows the various ways the voltage $V$, the magnetic field $B$, and the 
temperature difference $\Delta T$ enter the transport characteristics.

However, one should note that within the EOM method not only many of the leading, but also
some very-high-order contributions---in the sense of perturbation theory---are included, hence an interpretation of
the results in terms of low-order processes (which can be quite illuminating, if applicable, see, {\it e.g.},
Ref.~\cite{gergs2015}) is beyond reach.

In order to elucidate the role of correlated hopping, and find possible ways to infer its existence 
(and maybe even its value) from transport experiments, we studied in detail the spectral and transport properties of the QD system. The data presented in Fig.~(\ref{fig:rys11})
show that the model for $x=0$ leads to a conductance symmetric and a thermopower anti-symmetric 
with respect to $\delta=0$. A sizeable value of $x$ implies a distinct asymmetry, namely 
different heights of the conductance peaks and clear changes in the thermopower. Thus we conclude that 
a thorough symmetry analysis of the consecutive peaks in the conductance, and of the related features in the thermopower,
can provide information on the very existence of the correlated hopping term as well as its magnitude.
One of the conductance peaks of the single-level quantum dot is not affected by a change in $x$, while the other 
decreases in height. The increase of $[G(0)-G(x)]/G(0)$, where $G(0)$ is the conductance maximum 
at the upper peak and $G(x)$ its value at the lower peak, is faster than linear with $x$, 
{\it i.e.}, faster than the function $y=a\cdot x$ ($a=0.04$ for the parameters in Fig.~(\ref{fig:rys10_})).  
Measuring the peak variation and calculating the value of $a$ for the known parameters of the experimental
setup hence allows for a conservative estimation of $x$ and thus the correlated-hopping term. Note that
$\Gamma^L$ also can be estimated from experiment, as it is approximately given by the half-width of the upper conductance peak. 

The anisotropy of the couplings affects the conductance and thermopower, as shown in Figs.~(\ref{fig:rys7}a)
and (\ref{fig:rys7}b), respectively. This figure exhibits the transport coefficients for a few values 
of the asymmetry ({\it i.e.}, $\Gamma^R/\Gamma^L$) for $x=0.5$. An increase of this parameter 
mainly affects the half-width and the height of one of the peaks. 
In our setup, the impaired peak corresponds to large $\delta$.
However, the anisotropy may mask the effect of the $x$ parameter,
thus making its unique identification uncertain. However, as discussed above 
for a strongly 
asymmetric coupling the simultaneous measurement of the gate bias dependence of the thermopower
provides additional information from which the very existence of a non-zero $x$ can be inferred.

The non-equilibrium Green function technique in conjunction with the EOM allows the study of
transport coefficients beyond the linear approximation, as exemplified above.
In this paper, we have limited ourselves to calculations of the differential conductance:
the dependence of $G_d$ on $x$ and $\delta$ for a number of voltages is shown in Sec.~\ref{sec:nl}. 
The departures from the linear (small voltage) results depend on $x$ and $\delta$ in a complicated way. 
However, for 
the studied parameter values the conductance decreases with increasing $V$, except in the limits of 
a nearly empty or doubly occupied dot, {\it i.e.}, at the outer wings of the conductance peaks.
This is well visible in panel (b) of Fig.~(\ref{fig:rys12}) for $\delta\lesssim -6$ 
and $\delta \gtrsim 6$.      

\acknowledgments{
The work reported here has been supported by the M.~Curie-Sk\l{}odowska University,
National Science Center grant DEC-2017/27/B/ST3/01911 (Poland), the
Deutsche Forschungsgemeinschaft (project number 107745057, TRR 80),
and the University of Augsburg.}

\appendix
\section{Charge and heat currents}
\label{app:curr}
The charge current out of the electrode $\lambda$ is calculated
as time derivative of the average charge in that electrode~\cite{meir1992,meir1994},
$\langle N_{\lambda}\rangle=\sum_{k\sigma} \langle n_{\lambda{k}\sigma}\rangle$:
\be
I_{\lambda}=-e\left\langle\frac{dN_{\lambda}}{dt}\right\rangle=\frac{ie}{\hbar}\langle [N_\lambda,\hat{H}]\rangle
\label{current1}
\ee
where $\langle...\rangle$ denotes the statistical average.
The calculation of the heat flux follows that of the charge. The heat flux is
\be
J_\lambda=\frac{i}{\hbar}\langle [H_\lambda,\hat{H}]\rangle -\mu_\lambda\frac{i}{\hbar}\langle [N_\lambda,\hat{H}]\rangle,
\label{hcurrent}
\ee 
where
$H_\lambda=\sum_{{k} \sigma}\varepsilon_{\lambda {k} \sigma}n_{\lambda {k} \sigma}$ 
is the energy operator for the electrode $\lambda$. 
Calculating the commutators and introducing appropriate GFs, one finds
\beq
I_{\lambda}(t)&=&\frac{2e}{\hbar}\sum_{{k}\sigma}{\rm Re}
\bigg[{V}_{\lambda{k}\sigma}G^{<}_{\sigma,\lambda{k}\sigma}(t,t)\bigg],
\label{curr1} \\
J_{\lambda}(t)&=&\frac{2e}{\hbar}\sum_{{k}\sigma}(\varepsilon_{\lambda {k}}-\mu_\lambda) {\rm Re}
\bigg[{V}_{\lambda{k}\sigma}G^{<}_{\sigma,\lambda{k}\sigma}(t,t)\bigg].
\label{hcurr1}
\eeq
Here the GF $G^{<}_{\sigma,\lambda{k}\sigma}(t,t')=i\langle c^\dagger_{\lambda k\sigma}(t')D_\sigma(t)\rangle$ denotes
the lesser GF.
Using the standard approach \cite{haug-jauho1996} to calculate time-ordered functions, one obtains
the final expressions for the stationary currents in the following general form:
\beq 
I_{\lambda}&=&\frac{ie}{\hbar} \int\frac{dE}{2\pi}\sum_\sigma \Gamma_\sigma^{\lambda}(E)
\{ G_\sigma^{<}(E) \nonumber \\
&+& f_{\lambda}(E) [G_\sigma^{r}(E)-G_\sigma^{a}(E)]\}, 
\label{charge-curr} \\
J_{\lambda}&=&\frac{ie}{\hbar} \int\frac{dE}{2\pi}\sum_\sigma\Gamma_\sigma^{\lambda}(E)
(E-\mu_\lambda)\{ G_\sigma^{<}(E) \nonumber \\
&+& f_{\lambda}(E) [G_\sigma^{r}(E)-G_\sigma^{a}(E)]\}.
\label{heat-curr}
\eeq
The parameters $\Gamma_\sigma^{\lambda}(E)=2\pi\sum_{{k}}|V_{\lambda{k}\sigma}|^2\delta(E-\varepsilon_{\lambda{k}})$
describe the coupling between the dot and the electrode, and we write the equations for the stationary currents
{\it via} Fourier transforms of the GFs $G^{i}_\sigma(E)=\langle\langle D_\sigma|D^\dagger_\sigma\rangle\rangle^i_{E}$ 
with $i=r,a,<$ denoting retarded, advanced, and lesser functions.
Since the GFs $G^{i}_\sigma(E)=\langle \langle  D_\sigma|D^\dagger_\sigma\rangle\rangle^i_{E}$  
determine the transport properties of the system, we call them transport GFs in the following. 
On the other hand, it is important to note that the spectral properties of the dot (like the density of states) 
are given by another GF, defined with the operators $d_\sigma$ and $d^\dagger_\sigma$, 
namely $g_\sigma(E)=\langle  \langle  d_\sigma|d^\dagger_\sigma\rangle\rangle_{E}$.
For example, the dot spectral function $A_\sigma(E)$ at energy $E$ is 
given as $A_\sigma(E)=- \mathrm{Im} g_\sigma(E+i0)/\pi$, and the equilibrium charge density
$\langle n_\sigma \rangle$ equals the integral $\int dE A_\sigma(E)f(E)$, 
where $f(E)$ is the Fermi distribution at temperature $T$ and chemical potential $\mu$. 
 
Having in mind non-equilibrium charge and heat transport induced by a voltage or a temperature difference
across the system, we keep the dependence of the Fermi distribution functions $f_\lambda(E)$
on the electrode at hand {\it via} its chemical potential $\mu_\lambda$ and temperature $T_\lambda$. 
The heat current (\ref{heat-curr}) can be written as difference between
the energy current $J^E_\lambda$ and the charge current $I_\lambda$:
\be
J_\lambda=J^E_\lambda-\mu_\lambda I_\lambda.
\ee
The standard application of the EOM technique gives retarded and advanced GFs. To calculate the currents, one 
also needs the lesser GF $G_\sigma^<(E)$ entering (\ref{charge-curr}) and (\ref{heat-curr}).
In the literature various proposals have been used to obtain this function, some of them 
relying on the approximate calculations, others making use of 
proportionate couplings \cite{meir1992} $\Gamma_\sigma^L(E)=\alpha \Gamma_\sigma^R(E)$ with 
$\alpha =$ const. Here we shall present an expression which relates the transport lesser GF exactly to 
its retarded and advanced counterparts; the relation is exact in the wide-band limit.
In this limit, the effective couplings $\Gamma^\lambda_\sigma (E)=\Gamma^\lambda_\sigma$ 
do not depend on energy, and one finds (see the Suppl. Material \cite{sm} for details):
\beq  
 \langle D^\dagger_\sigma D_\sigma\rangle&=& -i\int \frac{dE}{2\pi} G^<(E) 
 \label{DD-noneq1} \\
&=&i\int \frac{dE}{2\pi}\frac{\sum_\lambda \Gamma_\sigma^\lambda f_\lambda(E)}{\sum_\lambda\Gamma_\sigma^\lambda}
[G^r_\sigma(E)-G^a_\sigma(E)]. \nonumber
\eeq
This sum rule for the correlated-hopping model, which is exact in the wide-band limit, 
is an important formal result of our paper. 
Its proof is given in the Suppl. Material \cite{sm}. The sum rule (\ref{DD-noneq1}) extends 
that found earlier \cite{lavagna2015} for the standard single-impurity Anderson model. 

Using the above result for the lesser GF, one finds the currents flowing out of 
the $\lambda$ electrode as follows:
\beq \label{charge-curr-la}
I_{\lambda}&=&\frac{2e}{\hbar} \int\frac{dE}{2\pi}\sum_\sigma \Gamma_\sigma^{\lambda}
\nonumber \\
&\times& \frac{\sum_{\lambda'} \Gamma^{\lambda'}_\sigma(f_{\lambda'}(E)-f_{\lambda}(E))}{\sum_{\lambda'} \Gamma^{\lambda'}_\sigma} 
\mathrm{Im} G_\sigma^{r}(E), \\
J_{\lambda}&=&\frac{2e}{\hbar} \int\frac{dE}{2\pi}\sum_\sigma\Gamma_\sigma^{\lambda}(E)
(E-\mu_\lambda)  \nonumber \\
&\times& \frac{\sum_{\lambda'} \Gamma^{\lambda'}_\sigma(f_{\lambda'}(E)-f_{\lambda}(E))}{\sum_{\lambda'} \Gamma^{\lambda'}_\sigma} 
\mathrm{Im} G_\sigma^{r}(E).
\label{heat-curr-la}
\eeq
These expressions can be used for calculating the currents in an arbitrary system consisting 
of the central dot and several terminals.

Formally the above manipulations are similar to those arising in the calculation of the currents in the standard Anderson 
model \cite{meir1994}. However, here we deal with completely different GFs. Moreover, as we shall 
see below, to calculate the transport GF one also needs the spectral one. Note that the kinetic and transport 
coefficients are expressed through the imaginary part of the transport GF only. Due to this fact,
we shall denote the imaginary part of the transport GF as ``transport density of states''.

\section{Calculation of the transport Green function}
\label{app:trGF}
Before calculating the relevant GF, let us note the following identities:
\beq
D_{\bar{\sigma}} D_{\sigma}&=&(1-x)d_{\bar{\sigma}} d_{\sigma}, \\
n_{\bar{\sigma}} D_{\sigma}&=&(1-x)n_{\bar{\sigma}} d_{\sigma},
\label{nd-nD} \\
D_{\bar{\sigma}} n_{\sigma}&=&(1-x)d_{\bar{\sigma}} n_{\sigma},\\
D^{\dagger}_{\bar{\sigma}}c_{\lambda k\bar{\sigma}} D_{\sigma}
&=&d^{\dagger}_{\bar{\sigma}}c_{\lambda k\bar{\sigma}} d_{\sigma},
\eeq
which will be occasionally used in various formulae below. The above identities
show that the point $x=1$ is a special one. Indeed, the model at hand is symmetric with respect to
$x=1$ for $0\le x\le 2$. For $x=0$ the only hopping is that of single-particle type $V_{\lambda k \sigma}$, 
while for $ n_{-\sigma}=1$ and $x=2$ one gets the effective hopping equal $-V_{\lambda k \sigma}$.
This together with the fact that $V_{\lambda k \sigma}$ enters all formulae as $|V_{\lambda k \sigma}|^2$
explains the equivalence of the model at these two limiting points. We remark in passing that a similar change of  
sign of the effective hybridization is also observed in the periodic Anderson model~\cite{wysokinski2014}. 
The symmetry of the present model goes beyond these two points, $x=0,2$, and is valid 
for arbitrary $x\in [0,2]$ as discussed earlier \cite{tooski2014}. 
It has to be stressed that within the present approach only the transport GF fulfils this symmetry, 
while the spectral one does not, as  discussed in Sec.~\ref{sec:res}.

To find the transport GF, we apply the EOM technique to two-time GFs and perform the 
appropriate decoupling. The quality of the solution
in this method depends on the decoupling procedure. Before proceeding, let us recall that 
the decouplings in the EOM technique typically are not well controlled.
We shall benchmark the proposed approximation scheme by checking the symmetry of the solution with respect to 
changing $x\leftrightarrow 2-x$. We start with the calculation of the transport GF, but  
as will be evident higher-order GFs are needed to solve
the system of equations. The coupling between various GFs is provided by some correlation functions, 
{\it inter alia} including the average occupation of the dot $\langle n_{\bar{\sigma}}\rangle$.

In Zubarev notation \cite{zubarev1960} for fermionic operators $A$ and $B$,
the equation for the two-time GF written in frequency $\omega$ space reads
\be
\omega\langle\langle A|B\rangle\rangle_\omega=\langle \{A,B\} \rangle +\langle\langle [A,H]|B\rangle\rangle_\omega.
\label{eom-freq}
\ee
Application of the above EOM to operators $A=D_\sigma$ and $B=D^\dagger_\sigma$ provides
\begin{widetext}
\beq
[\omega-\varepsilon_{\sigma}]\langle\langle D_{\sigma}|D^{\dagger}_{\sigma}\rangle\rangle_{\omega} = 
1 -x(2-x)\langle n_{\bar{\sigma}}\rangle +  \sum_{\lambda k}V^{*}_{\lambda k \sigma}\langle\langle c_{\lambda k \sigma} |D^{\dagger}_{\sigma}\rangle\rangle_{\omega}+U\langle\langle n_{\bar{\sigma}}D_{\sigma}|D^{\dagger}_{\sigma}\rangle\rangle_{\omega} 
\nonumber \\
-x(2-x)\sum_{\lambda k}\left[V^*_{\lambda k {\sigma}}\langle\langle n_{\bar{\sigma}}c_{\lambda k \sigma} 
| D^{\dagger}_{\sigma}\rangle\rangle_{\omega}  
 + V^{*}_{\lambda k \bar{\sigma}}\langle\langle D^{\dagger}_{\bar{\sigma}} c_{\lambda k\bar{\sigma}} D_{\sigma} 
|D^{\dagger}_{\sigma}\rangle\rangle_{\omega}\right]. 
\label{(A-g-t)}
\eeq
The EOM for the next GF,
\beq
(\omega-\varepsilon_{\lambda k})\langle\langle c_{\lambda k \sigma} |D^{\dagger}_{\sigma}\rangle\rangle_{\omega}=
V_{\lambda k \sigma}\langle\langle D_{\sigma}|D^{\dagger}_{\sigma}\rangle\rangle_{\omega} ,
\label{eq:D3-t}
\eeq
allows one to write 
\be
\sum_{\lambda k}V^{*}_{\lambda k \sigma}\langle\langle c_{\lambda k \sigma} |D^{\dagger}_{\sigma}\rangle\rangle_{\omega}=\Sigma_{0\sigma}(\omega)\langle\langle D_{\sigma}|D^{\dagger}_{\sigma}\rangle\rangle_{\omega}.
\ee
The factor in front of the GF on the rhs of the preceding equation defines the self-energy:
\beq
\Sigma_{0\sigma} (\omega)=\sum_{\lambda k} \frac{|V_{\lambda k \sigma}|^{2}}{\omega-\varepsilon_{\lambda k}}.
\label{sigma0}
\eeq
In the wide-band limit one approximates (\ref{sigma0}) by its imaginary part:
\be
\Sigma_{0\sigma} (\omega)\approx -i\pi \sum_{\lambda k}{|V_{\lambda k \sigma}|^{2}}\delta(\omega-\varepsilon_{\lambda k}) 
=-i\frac{1}{2}\sum_\lambda \Gamma_\sigma^\lambda(\omega)=-i(\Gamma^L_\sigma+\Gamma^R_\sigma)/2
=-i\bar{\Gamma}_\sigma/2,
\label{Gam-wbl}
\ee
typically assumed to be energy independent.
The higher-order GF which multiplies $U$ in Eq.~(\ref{(A-g-t)}) reads
\beq
[\omega-\varepsilon_{\sigma}-U]\langle\langle n_{\bar{\sigma}}D_{\sigma} |D^{\dagger}_{\sigma}\rangle\rangle _{\omega}= (1-x)^2\langle n_{\bar{\sigma}}\rangle 
- \sum_{\lambda k} V_{\lambda k \bar{\sigma}}\langle\langle c^{\dagger}_{\lambda k \bar{\sigma}} D_{\bar{\sigma}} 
D_{\sigma} |D^{\dagger}_{\sigma}\rangle\rangle_{\omega} 
\nonumber \\ 
+(1-x)^2 \sum_{\lambda k} \left[V^{*}_{\lambda k\sigma}\langle\langle n_{\bar{\sigma}} c_{\lambda k \sigma} 
|D^{\dagger}_{\sigma}\rangle\rangle_{\omega} 
+  V^{*}_{\lambda k \bar{\sigma}}\langle\langle D^{\dagger}_{\bar{\sigma}} c_{\lambda k \bar{\sigma}} D_{\sigma} | 
D^{\dagger}_{\sigma}\rangle\rangle_{\omega}\right].
\label{(D-g-t)}
\eeq
Equation (\ref{(D-g-t)}) suggests that the calculation of the transport GF, independently of 
the forthcoming decouplings, requires the knowledge of $\langle n_{\bar{\sigma}}\rangle$ and 
thus of the spectral GF, $\langle\langle d_{\bar{\sigma}} |d^{\dagger}_{\bar{\sigma}}\rangle\rangle _{\omega}$.
The remaining GFs entering the rhs of Eqs.~(\ref{(A-g-t)}) and (\ref{(D-g-t)}) fulfil 
\beq
[\omega-\varepsilon_{\lambda k}]\langle\langle n_{\bar{\sigma}}c_{\lambda k\sigma}|D^{\dagger}_{\sigma}\rangle\rangle_{\omega}= 
V_{\lambda k\sigma}\langle\langle n_{\bar{\sigma}} D_{\sigma}|D^{\dagger}_{\sigma}\rangle\rangle_{\omega}
-\sum_{\lambda' k'} V_{\lambda' k'\bar{\sigma}}\langle\langle c^{\dagger}_{\lambda' k'\bar{\sigma}}
 D_{\bar{\sigma}} c_{\lambda k \sigma}|D^{\dagger}_{\sigma}\rangle\rangle_{\omega}  \nonumber \\
+\sum_{\lambda'k'}V^{*}_{\lambda'k'\bar{\sigma}}\langle\langle D^{\dagger}_{\bar{\sigma}} c_{\lambda' k'\bar{\sigma}}
c_{\lambda k \sigma}|D^{\dagger}_{\sigma}\rangle\rangle_{\omega},
\label{(B-g-t)}
\eeq
\beq
[\omega -\varepsilon_{\lambda k}-\varepsilon_{\sigma} +\varepsilon_{\bar{\sigma}}]\langle\langle 
D^{\dagger}_{\bar{\sigma}} c_{\lambda k \bar{\sigma}} D_{\sigma}| D^{\dagger}_{\sigma} \rangle\rangle _{\omega} =
\langle D^{\dagger}_{\bar{\sigma}} c_{\lambda k \bar{\sigma}}\rangle + 
 V_{\lambda k \bar{\sigma}}\langle\langle n_{\bar{\sigma}} 
D_{\sigma}|D^{\dagger}_{\sigma}\rangle\rangle _{\omega}  \nonumber \\
- \sum_{\lambda'k'} V_{\lambda' k' \bar{\sigma}}\langle\langle 
c^{\dagger}_{\lambda' k' \bar{\sigma}}c_{\lambda k \bar{\sigma}} D_{\sigma}|D^{\dagger}_{\sigma}\rangle\rangle _{\omega} 
+\sum_{\lambda'k'} V^{*}_{\lambda'k'\sigma}\langle\langle D^{\dagger}_{\bar{\sigma}} c_{\lambda k \bar{\sigma}} 
c_{\lambda' k'\sigma}|D^{\dagger}_{\sigma}\rangle\rangle _{\omega},
\label{(C-g-t)}
\eeq
\beq
[\omega+\varepsilon_{\lambda k}-\varepsilon_{\sigma}-\varepsilon_{\bar{\sigma}}-U]\langle\langle c^{\dagger}_{\lambda k \bar{\sigma}} D_{\bar{\sigma}} 
D_{\sigma} | D^{\dagger}_{\sigma}\rangle\rangle_{\omega}= 
(1-x)^2 \langle c^{\dagger}_{\lambda k \bar{\sigma}} D_{\bar{\sigma}}\rangle +x(2-x)\langle c^{\dagger}_{\lambda k \bar{\sigma}} D_{\bar{\sigma}}n_\sigma\rangle
-V^{*}_{\lambda k \bar{\sigma}}\langle\langle n_{\bar{\sigma}} D_{\sigma} |D^{\dagger}_{\sigma}\rangle\rangle_{\omega} \nonumber \\
+(1-x)^2 \sum_{\lambda' k'}\left[V^{*}_{\lambda' k'\sigma}\langle\langle c^{\dagger}_{\lambda k \bar{\sigma}} d_{\bar{\sigma}} 
c_{\lambda' k' \sigma} |D^{\dagger}_{\sigma}\rangle\rangle_{\omega}   
+V^{*}_{\lambda' k' \bar{\sigma}} \langle\langle c^{\dagger}_{\lambda k \bar{\sigma}} c_{\lambda' k' \bar{\sigma}} 
d_{\sigma} |D^{\dagger}_{\sigma}\rangle\rangle _{\omega}\right] \nonumber \\
+ x \sum_{\lambda' k'}\left[V^{*}_{\lambda' k' \sigma}\langle\langle c^{\dagger}_{\lambda k \bar{\sigma}} D_{\bar{\sigma}} n_{\sigma} 
c_{\lambda' k' \sigma} |D^{\dagger}_{\sigma}\rangle\rangle_{\omega}  
+V^{*}_{\lambda' k' \bar{\sigma}} \langle\langle 
c^{\dagger}_{\lambda k \bar{\sigma}} c_{\lambda' k' \bar{\sigma}}n_{\bar{\sigma}} D_{\sigma}|D^{\dagger}_{\sigma}\rangle\rangle_{\omega}\right].
\label{(E-g-t)}
\eeq
With the auxiliary notation
\beq
S^{\mathrm{tr}}_n&=&\sum_{\lambda k}V^{*}_{\lambda k \sigma}\langle\langle n_{\bar{\sigma}} c_{\lambda k \sigma} |D^{\dagger}_{\sigma}\rangle\rangle_{\omega} \\
S^{\mathrm{tr}}_d&=&\sum_{\lambda k}V^{*}_{\lambda k \bar{\sigma}}\langle\langle D^{\dagger}_{\bar{\sigma}} c_{\lambda k\bar{\sigma}} D_{\sigma} |D^{\dagger}_{\sigma}\rangle\rangle_{\omega} \\
S^{\mathrm{tr}}_c&=&\sum_{\lambda k}V_{\lambda k \bar{\sigma}}\langle\langle c^{\dagger}_{\lambda k \bar{\sigma}} D_{\bar{\sigma}} 
D_{\sigma} | D^{\dagger}_{\sigma}\rangle\rangle_{\omega}
\eeq
one finds   
\be 
[\omega-\varepsilon_{\sigma}-\Sigma_{0 \sigma}]\langle\langle D_{\sigma} |D^{\dagger}_{\sigma}\rangle\rangle _{\omega}= 
1 - x(2-x)\langle n_{\bar{\sigma}}\rangle - x(2-x)(S^{\mathrm{tr}}_n+S^{\mathrm{tr}}_d) + U\langle\langle n_{\bar{\sigma}}D_{\sigma}|D^{\dagger}_{\sigma}\rangle\rangle_{\omega}
\label{(D-t)}
\ee
\be 
[\omega-\varepsilon_{\sigma}-U]\langle\langle n_{\bar{\sigma}}D_{\sigma} |D^{\dagger}_{\sigma}\rangle\rangle _{\omega}= 
(1-x)^2\langle n_{\bar{\sigma}}\rangle -S^{\mathrm{tr}}_c+(1-x)^2(S_n^{\mathrm{tr}}+S_d^{\mathrm{tr}}).  
\label{(nD-t)}
\ee
\end{widetext}
We shall not calculate the GFs containing two 
$c_{\lambda k \sigma}$ operators but approximate the GFs in question, avoiding the
appearance of functions which describe spin-flip processes. Thus we project higher-order GFs as follows: 
\be
\langle\langle  c^{\dagger}_{\lambda'k' \bar{\sigma}} c_{\lambda k\bar{\sigma}}D_{\sigma}|D^{\dagger}_{\sigma}\rangle\rangle _{\omega}\approx 
\langle c^{\dagger}_{\lambda' k'\bar{\sigma}} c_{\lambda k \bar{\sigma}} \rangle\langle\langle  D_{\sigma}|D^{\dagger}_{\sigma}\rangle\rangle _{\omega},
\label{proj-start}
\ee
and 
\be
\langle\langle  c^{\dagger}_{\lambda'k' \bar{\sigma}} D_{\bar{\sigma}}c_{\lambda k\sigma}|D^{\dagger}_{\sigma}\rangle\rangle _{\omega}\approx 
\langle c^{\dagger}_{\lambda'k'\bar{\sigma}}D_{\bar{\sigma}} \rangle\langle\langle c_{\lambda k \sigma}|D^{\dagger}_{\sigma}\rangle\rangle _{\omega},
\ee
\be
\langle\langle  c^{\dagger}_{\lambda k \bar{\sigma}} d_{\bar{\sigma}}c_{\lambda' k'\sigma}|D^{\dagger}_{\sigma}\rangle\rangle _{\omega}\approx 
\langle c^{\dagger}_{\lambda k\bar{\sigma}}d_{\bar{\sigma}} \rangle\langle\langle c_{\lambda' k' \sigma}|D^{\dagger}_{\sigma}\rangle\rangle _{\omega},
\ee
\be
\langle\langle  D^{\dagger}_{\bar{\sigma}} c_{\lambda' k'\bar{\sigma}} c_{\lambda k\sigma}|D^{\dagger}_{\sigma}\rangle\rangle _{\omega} \approx 
\langle D^{\dagger}_{\bar{\sigma}} c_{\lambda' k' \bar{\sigma}} \rangle\langle\langle c_{\lambda k\sigma}|D^{\dagger}_{\sigma}\rangle\rangle _{\omega},
\ee
\be
\langle\langle  D^{\dagger}_{\bar{\sigma}} c_{\lambda k\bar{\sigma}} c_{\lambda' k'\sigma}|D^{\dagger}_{\sigma}\rangle\rangle _{\omega} \approx 
\langle D^{\dagger}_{\bar{\sigma}} c_{\lambda k \bar{\sigma}} \rangle\langle\langle c_{\lambda' k'\sigma}|D^{\dagger}_{\sigma}\rangle\rangle _{\omega}.
\ee
As already alluded to, the above decouplings are analogous to those which in the context of standard Hubbard 
model are known as Lacroix decouplings~\cite{lacroix1981}. The decoupling of the following GF,
\be
\langle\langle  c^{\dagger}_{\lambda'k' \bar{\sigma}} c_{\lambda k\bar{\sigma}}d_{\sigma}|D^{\dagger}_{\sigma}\rangle\rangle _{\omega}\approx 
\langle c^{\dagger}_{\lambda' k'\bar{\sigma}} c_{\lambda k \bar{\sigma}} \rangle\langle\langle  d_{\sigma}|D^{\dagger}_{\sigma}\rangle\rangle _{\omega},
\label{proj-fin}
\ee
introduces a novel GF which has not appeared hitherto, namely 
$\langle\langle d_{\sigma}|D^{\dagger}_{\sigma}\rangle\rangle _{\omega}$. To obtain this one, we use an exact relation:
\be
\langle \langle  D_\sigma|D^\dagger_\sigma\rangle\rangle_{\omega}=\langle \langle  d_\sigma|D^\dagger_\sigma\rangle\rangle_{\omega}
-x \langle \langle  d_\sigma n_{\bar{\sigma}}|D^\dagger_\sigma\rangle\rangle_{\omega},
\label{equiv-DD}
\ee
deduced from the operator identity $D_\sigma=d_\sigma-xd_\sigma n_{\bar{\sigma}}$. 
If the GF at hand is multiplied by $(1-x)$, we can express it by the functions appearing on the 
lhs of Eqs.~(\ref{(D-t)}) and (\ref{(nD-t)}):
\be
(1-x)\langle \langle  d_\sigma|D^\dagger_\sigma\rangle\rangle_{\omega}
=(1-x)\langle \langle  D_\sigma|D^\dagger_\sigma\rangle\rangle_{\omega}
+x \langle \langle n_{\bar{\sigma}} D_\sigma |D^\dagger_\sigma\rangle\rangle_{\omega}.
\label{equiv-DD2}
\ee
This closes the system of equations for $\langle\langle D_{\sigma} |D^{\dagger}_{\sigma}\rangle\rangle _{\omega}$, except 
that we still need the spectral GF to calculate $\langle n_{\bar{\sigma}}\rangle$.
It is worth noting in advance that the spectral GF 
$\langle\langle d_{\sigma} |d^{\dagger}_{\sigma}\rangle\rangle _{\omega}$ turns out to be coupled back to the
transport one and the function $\langle\langle n_{\bar{\sigma}}d_{\sigma} |D^{\dagger}_{\sigma}\rangle\rangle _{\omega}$. 

The solution of (\ref{(D-t)}) and (\ref{(nD-t)}) is a relatively easy task. First, using the presented 
decouplings, one calculates the parameters $S^{\mathrm{tr}}_n$, $S^{\mathrm{tr}}_d$, and $S^{\mathrm{tr}}_c$. 
For $S^{\mathrm{tr}}_n$ one finds
\beq
S^{\mathrm{tr}}_n&=&\Sigma_{0 \sigma}\langle\langle n_{\bar{\sigma}}D_\sigma |D^\dagger_\sigma\rangle\rangle_{\omega} \nonumber \\
&+&\sum_{\lambda' k'\bar{\sigma}}\left[V^{*}_{\lambda' k'\bar{\sigma}}\langle D^\dagger c_{\lambda' k'\bar{\sigma}}\rangle
-V^{*}_{\lambda' k'\bar{\sigma}}\langle c^\dagger_{\lambda' k'\bar{\sigma}}D_{\bar{\sigma}}\rangle\right] \nonumber \\ 
&\times &\Sigma^{\prime}_\sigma (\omega) 
\langle\langle D_\sigma |D^\dagger_\sigma\rangle\rangle_{\omega},
\eeq  
where 
\be
\Sigma^{\prime}_\sigma(\omega)=\sum_{\lambda k}\frac{|V_{\lambda k \sigma}|^2}{(\omega-\varepsilon_{\lambda k})^2},
\ee
vanishes in the wide-band limit and for energy independent coupling, the approximation assumed to be valid here.
Thus we end up with
\be
S^{\mathrm{tr}}_n=\Sigma_{0 \sigma}\langle\langle n_{\bar{\sigma}}D_\sigma |D^\dagger_\sigma\rangle\rangle_{\omega},
\label{sn-tr}
\ee
where we have omitted the frequency dependence of the self-energy $\Sigma_{0\sigma}(\omega)$. Occasionally we
shall use this convention in the following. Let us note that the proposed decouplings of the GFs 
containing two operators describing the electrons on the leads provide  
a simple expressions for $S^{\mathrm{tr}}_{n,d,c}$ in terms of   
$\langle\langle D_\sigma |D^\dagger_\sigma\rangle\rangle_{\omega}$ and 
$\langle\langle n_{\bar{\sigma}}D_\sigma |D^\dagger_\sigma\rangle\rangle_{\omega}$ only. Later on we shall
apply analogous decouplings to find the spectral GFs $\langle\langle d_\sigma |d^\dagger_\sigma\rangle\rangle_{\omega}$
and $\langle\langle n_{\bar{\sigma}}d_\sigma |d^\dagger_\sigma\rangle\rangle_{\omega}$.

The remaining two auxiliary parameters read
\beq
\label{sd-tr}
S^{\mathrm{tr}}_d&=&\tilde{b}_{1\bar{\sigma}}+ \Sigma^{(1)}_{\bar{\sigma}}\langle\langle n_{\bar{\sigma}}D_\sigma |D^\dagger_\sigma\rangle\rangle_{\omega} \nonumber \\
&+& \left[\tilde{b}_{1\bar{\sigma}}\Sigma_{0\sigma}-\Sigma^{T}_{1\bar{\sigma}}\right]\langle\langle D_\sigma |D^\dagger_\sigma\rangle\rangle_{\omega}, \\
S^{\mathrm{tr}}_c &=& \bar{b}_{2\bar{\sigma}}
+\left[(1-x)^2\Sigma^T_{2\bar{\sigma}}+\bar{b}_{2\bar{\sigma}}\Sigma_{0\sigma}\right]\langle\langle D_\sigma |D^\dagger_\sigma\rangle\rangle_{\omega}
\nonumber \\
&+&  \left[x(2-x)\Sigma^T_{2\bar{\sigma}} -\Sigma^{(2)}_{\bar{\sigma}}\right]\langle\langle n_{\bar{\sigma}}D_\sigma |D^\dagger_\sigma\rangle\rangle_{\omega},
\label{sc-tr}
\eeq 
where the novel symbols denote the various ``summed'' correlation functions or self-energies. For example,
\be
\tilde{b}_{1\bar{\sigma}} (\omega)=\sum_{\lambda k}\frac{V^{*}_{\lambda k \bar{\sigma}}
\langle D^{\dagger}_{\bar{\sigma}} c_{\lambda k \bar{\sigma}}\rangle}{\omega-\varepsilon_{\lambda k}-\varepsilon_1 +i\tilde{\gamma}^{\bar{\sigma}}_1},
\label{deftbibs}
\ee
and 
\be
\tilde{b}_{2\bar{\sigma}} (\omega)=\sum_{\lambda k}\frac{V_{\lambda k \bar{\sigma}}
\langle  c^{\dagger}_{\lambda k \bar{\sigma}}D_{\bar{\sigma}}\rangle}{\omega+\varepsilon_{\lambda k}-\varepsilon_2 +i\tilde{\gamma}_2}.
\label{deftbibs2}
\ee
Using the definition of the operator $D_\sigma$, the last correlation function can be split 
as, {\it e.g.}, $\tilde{b}_{1\bar{\sigma}} (\omega)={b}_{1\bar{\sigma}} (\omega)-x{N}_{1\bar{\sigma}} (\omega)$, with
obvious definitions of ${b}_{1\bar{\sigma}} (\omega)$ and ${N}_{1\bar{\sigma}} (\omega)$. 

The other self-energies entering (\ref{sd-tr}) and (\ref{sc-tr}) are defined as 
\be
\Sigma^{T}_{1\bar{\sigma}}(\omega)
=\sum_{\lambda k} \sum_{\lambda' k'} \frac{V^{*}_{\lambda k \bar{\sigma}} V_{\lambda' k' \bar{\sigma}} 
\langle c^{\dagger}_{\lambda' k' \bar{\sigma} } c_{\lambda k \bar{\sigma}}\rangle}{\omega-\varepsilon_{\lambda k} -\varepsilon_1 +i\tilde{\gamma}^{\bar{\sigma}}_1},
\ee
\be
\Sigma^{T}_{2\bar{\sigma}}(\omega)=\sum_{\lambda k}\sum_{\lambda' k'} \frac{V_{\lambda k\bar{\sigma}} V^{*}_{\lambda' k'\bar{\sigma}}
 \langle c^{\dagger}_{\lambda k\bar{\sigma}} c_{\lambda' k' \bar{\sigma}}\rangle}
{\omega+\varepsilon_{\lambda k}-\varepsilon_2 +i\tilde{\gamma}_2},
\ee
\be
\Sigma^{(1)}_{\bar{\sigma}}=\sum_{\lambda k}\frac{|V_{\lambda k\bar{\sigma}}|^{2}}{\omega-\varepsilon_{\lambda k}-\varepsilon_{1}+i\tilde{\gamma}^{\bar{\sigma}}_1},
\ee
\be
\Sigma^{(2)}_{\bar{\sigma}}=\sum_{\lambda k}\frac{|V_{\lambda k\bar{\sigma}}|^{2}}{\omega+\varepsilon_{\lambda k}-\varepsilon_{2}+i\tilde{\gamma}^{\bar{\sigma}}_1},
\ee
where the energies $\varepsilon_{1(2)}$ are defined as 
\beq
\varepsilon_{1}=\varepsilon_{\sigma}-\varepsilon_{\bar{\sigma}}
\label{eq:e1} \\
\varepsilon_{2}=\varepsilon_{\sigma}+\varepsilon_{\bar{\sigma}}+U.
\label{eq:e2}
\eeq
The symbols $\tilde{\gamma}^{\bar{\sigma}}_{1}$ and $\tilde{\gamma}_{2}$ refer to the inverse lifetimes of the 
singly (doubly) occupied states on the dot. For the standard Anderson model
they have been found \cite{lavagna2015} to play an decisive role in assuring the proper Kondo behavior 
at low temperatures. They are also important here due to the same reasons.

The correlation function $\bar{b}_{2\bar{\sigma}} (\omega)$ is given by the following combination:  
\beq 
\bar{b}_{2\bar{\sigma}} (\omega)=(1-x)[(1-x)b_{2\bar{\sigma}} (\omega)+xN_{2\bar{\sigma}}(\omega)] \nonumber \\
\equiv (1-x)\tilde{b}_{2\bar{\sigma}}.
\label{bar-tilde}
\eeq
Some of the self-energies are expressed by the transport and other by the spectral GF as shown in the Suppl. 
Material \cite{sm}, but to calculate $\bar{b}_{2\bar{\sigma}}(\omega)$ one needs   
$\langle\langle n_{\bar{\sigma}}D_\sigma|D^\dagger_\sigma\rangle\rangle_\omega$.   
On the other hand, the calculation of $\tilde{b}_{2\bar{\sigma}}$ 
requires the knowledge of the closely related function 
$\langle\langle n_{\bar{\sigma}}d_\sigma|D^\dagger_\sigma\rangle\rangle_\omega$. Thus the whole set of GFs 
required to solve the self-consistent set of equations comprises functions of diagonal:
$\langle\langle D_\sigma|D^\dagger_\sigma\rangle\rangle_\omega$, $\langle\langle d_\sigma|d^\dagger_\sigma\rangle\rangle_\omega$, and 
off-diagonal: $\langle\langle n_{\bar{\sigma}} d_\sigma|d^\dagger_\sigma\rangle\rangle_\omega$, 
$\langle\langle n_{\bar{\sigma}} D_\sigma|D^\dagger_\sigma\rangle\rangle_\omega$ character. 

The solution of Eqs.~(\ref{(D-t)}) and (\ref{(nD-t)}) for the transport GF  
is now written in a closed form. Defining the auxiliary function  
\begin{widetext}
\be
I_D(\omega)=\frac{U-x(2-x)(\Sigma_{0\sigma}+\Sigma^{(1)}_{\bar{\sigma}})}{\omega-\varepsilon_\sigma-U-(1-x)^2(\Sigma_{0\sigma}
+\Sigma^{(1)}_{\bar{\sigma}})-\Sigma_{\bar{\sigma}}^{(2)}+x(2-x)\Sigma^T_{2\bar{\sigma}}},
\label{sol-sp2-t}
\ee
which for $x=0$ reduces to $I_d$ ({\it cf.} Eq.~(\ref{sol-id}) in the main text)~\cite{eckern2020}, and finds 
\beq
\langle\langle D_\sigma|D^\dagger_\sigma\rangle\rangle_\omega=\frac{1-x(2-x)(\langle n_{\bar{\sigma}}\rangle+\tilde{b}_{1\bar{\sigma}})+n^D_{\mathrm{eff}}(\omega)I_D(\omega)}{\omega-\varepsilon_d-\Sigma_{0\sigma}+x(2-x)(\tilde{b}_{1\bar{\sigma}}\Sigma_{0\sigma}-
\Sigma^T_{1\bar{\sigma}})-I_D(\omega)B_D(\omega)},
\label{sol-gf-t}
\eeq
with 
\beq
n^D_{\mathrm{eff}}(\omega)&=&(1-x)^2(\langle n_{\bar{\sigma}}\rangle+\tilde{b}_{1\bar{\sigma}})-\bar{b}_{2\bar{\sigma}},
\label{sol-b44} \\
B_D(\omega)&=&(1-x)^2[\tilde{b}_{1\bar{\sigma}}\Sigma_{0\sigma} -\Sigma^T_{1\bar{\sigma}}-\Sigma^T_{2\bar{\sigma}}]
- \bar{b}_{2\bar{\sigma}}\Sigma_{0\sigma}, 
\label{sol-b45}
\eeq
and 
\be 
\Sigma_{ndD}(\omega)=(1-x)^2(\Sigma_{0\sigma}+\Sigma^{(1)}_{\bar{\sigma}}) 
- x(2-x)\Sigma^T_{2\bar{\sigma}}+\Sigma_{\bar{\sigma}}^{(2)}.
\label{sol-sp2-tD}
\ee
The related GF $\langle\langle n_{\bar{\sigma}}D_\sigma|D^\dagger_\sigma\rangle\rangle_\omega$ is given by
\be
\langle\langle n_{\bar{\sigma}}D_\sigma|D^\dagger_\sigma\rangle\rangle_\omega=\frac{(1-x)^2(\langle n_{\bar{\sigma}}\rangle+\tilde{b}_{1\bar{\sigma}})-\bar{b}_{2\bar{\sigma}} }{\omega-\varepsilon_\sigma-U-\Sigma_{ndD}} 
+\frac{[(1-x)^2(\tilde{b}_{1\bar{\sigma}}\Sigma_{0\sigma}-\Sigma^T_{1\bar{\sigma}}-\Sigma^T_{2\bar{\sigma}})
-\bar{b}_{2\bar{\sigma}}\Sigma_{0\sigma}}{\omega-\varepsilon_\sigma-U-\Sigma_{ndD}} \langle\langle D_\sigma|D^\dagger_\sigma\rangle\rangle_\omega.
\label{sol-gf-tn0-1}
\ee
Using the relations (\ref{nd-nD}) and (\ref{bar-tilde}), we get the last required GF:
\be
\langle\langle n_{\bar{\sigma}}d_\sigma|D^\dagger_\sigma\rangle\rangle_\omega=\frac{(1-x)(\langle n_{\bar{\sigma}}\rangle+\tilde{b}_{1\bar{\sigma}})-\tilde{b}_{2\bar{\sigma}} }{\omega-\varepsilon_\sigma-U-\Sigma_{ndD}} 
+\frac{[(1-x)(\tilde{b}_{1\bar{\sigma}}\Sigma_{0\sigma}-\Sigma^T_{1\bar{\sigma}}-\Sigma^T_{2\bar{\sigma}})
-\tilde{b}_{2\bar{\sigma}}\Sigma_{0\sigma}}{\omega-\varepsilon_\sigma-U-\Sigma_{ndD}} \langle\langle D_\sigma|D^\dagger_\sigma\rangle\rangle_\omega, 
\label{sol-gf-tn0-2}
\ee
\end{widetext}
The various symbols used above are 
summarised in App.~\ref{app:symbols}, where they are also expressed in terms  of the 
transport GFs and the auxiliary functions like (\ref{sol-gf-tn0-1}) and (\ref{sol-gf-tn0-2}). 
The calculation of the average occupation of the dot $\langle n_{\bar{\sigma}}\rangle$ requires the 
knowledge of the spectral GF (\ref{sol-gfd}). Also the self-energy $\tilde{b}_{1\bar{\sigma}}(\omega)$ 
requires the knowledge of the spectral GF and the related function 
$\langle\langle n_{\bar{\sigma}}d_\sigma|d^\dagger_\sigma\rangle\rangle$;  {\it cf.} Eq.~(\ref{b1bs-vs-sgf}).

\section{Self-energies in terms of the Green functions}
\label{app:symbols}
Here we list for completeness all self-energies entering the solutions 
expressed self-consistently in terms of the relevant GFs. 
The self-energy $\tilde{b}_{1\bar{\sigma}} (\omega)$ has been obtained in the previous section. 
It depends on the transport GF only: 
\beq
\tilde{b}_{1\bar{\sigma}} (\omega)=\sum_{\lambda k}\frac{V^{*}_{\lambda k \bar{\sigma}}
\langle D^{\dagger}_{\bar{\sigma}} c_{\lambda k \bar{\sigma}}\rangle}{\omega-\varepsilon_{\lambda k}-\varepsilon_1 +i\tilde{\gamma}^{\bar{\sigma}}_1} \nonumber \\
=\int \frac{d\varepsilon}{2\pi} \frac{\sum_{\lambda}\Gamma^{\lambda}_{\bar{\sigma}} f_{\lambda} (\varepsilon)\langle\langle D_{\bar{\sigma}} |D^{\dagger}_{\bar{\sigma}}\rangle\rangle ^{a}_{\varepsilon}}{\omega-\varepsilon 
-\varepsilon_1 +i\tilde{\gamma}^{\bar{\sigma}}_1}.
\eeq
The calculation of $\bar{b}_{2\bar{\sigma}} (\omega)$ requires the knowledge of ${b}_{2\bar{\sigma}} (\omega)$ 
and ${N}_{2\bar{\sigma}} (\omega)$, which are expressed as 
\beq
b_{2\bar{\sigma}}(\omega)
= \sum_{\lambda k}\frac{V_{\lambda k\bar{\sigma}}\langle c^{\dagger}_{\lambda k \bar{\sigma}}d_{\bar{\sigma}}\rangle}
{\omega+\varepsilon_{\lambda k}-\varepsilon_2+ i \tilde{\gamma}_2} \nonumber \\
= \int\frac{d\varepsilon}{2\pi} \frac{\sum_{\lambda} \Gamma^{\lambda}_{\bar{\sigma}} 
f_{\lambda}(\varepsilon) \langle\langle d_{\bar{\sigma}} |D^{\dagger}_{\bar{\sigma}}\rangle\rangle^{r}_{\varepsilon}}
{\omega +\varepsilon -\varepsilon_2+i\tilde{\gamma}_2}, 
\eeq
and
\beq
N_{2\bar{\sigma}} (\omega)=\sum_{\lambda k}\frac{V_{\lambda k \bar{\sigma}}
\langle c^{\dagger}_{\lambda  k \bar{\sigma}} d_{\bar{\sigma}} n_{\bar{\sigma}}\rangle}{\omega+ \varepsilon_{\lambda k} 
- \varepsilon _2 + i\tilde{\gamma}_2} \nonumber \\
=\int \frac{d\varepsilon}{2\pi} \frac{\sum_{\lambda}\Gamma^{\lambda}_{\bar{\sigma}} f_{\lambda} (\varepsilon)\langle\langle n_{\sigma} d_{\bar{\sigma}} |D^{\dagger}_{\bar{\sigma}}\rangle\rangle ^{r}_{\varepsilon}}{\omega+\varepsilon -\varepsilon_2 +i\tilde{\gamma}_2}
\eeq
At first glance the calculation of $b_{2\bar{\sigma}} (\omega)$ and $N_{2\bar{\sigma}} (\omega)$ requires
two new GFs. However, it turns out that these quantities enter the formulae for the
transport GFs in the combination $\bar{b}_{2\bar{\sigma}} (\omega)=(1-x)^2b_{2\bar{\sigma}} (\omega)
+x(1-x)N_{2\bar{\sigma}} (\omega)$. Using their definitions and the relation (\ref{equiv-DD}), one arrives at
\beq
\bar{b}_{2\bar{\sigma}} (\omega)=(1-x)^2\int \frac{d\varepsilon}{2\pi} \frac{\sum_{\lambda}\Gamma^{\lambda}_{\bar{\sigma}} f_{\lambda} (\varepsilon)\langle\langle D_{\bar{\sigma}} |D^{\dagger}_{\bar{\sigma}}\rangle\rangle ^{r}_{\varepsilon}}{\omega+\varepsilon -\varepsilon_2 +i\tilde{\gamma}_2} \nonumber \\
+x(2-x)\int \frac{d\varepsilon}{2\pi} \frac{\sum_{\lambda}\Gamma^{\lambda}_{\bar{\sigma}} f_{\lambda} (\varepsilon)\langle\langle n_{\sigma} D_{\bar{\sigma}} |D^{\dagger}_{\bar{\sigma}}\rangle\rangle ^{r}_{\varepsilon}}{\omega+\varepsilon -\varepsilon_2 +i\tilde{\gamma}_2}.
\label{b2bar}
\eeq
The remaining self-energies read 
\beq
\Sigma^{T}_{1\bar{\sigma}}(\omega)
=\sum_{\lambda k} \sum_{\lambda' k'} \frac{V^{*}_{\lambda k \bar{\sigma}} V_{\lambda' k' \bar{\sigma}} 
\langle c^{\dagger}_{\lambda' k' \bar{\sigma} } c_{\lambda k \bar{\sigma}}\rangle}{\omega-\varepsilon_{\lambda k} -\varepsilon_1 +i\tilde{\gamma}^{\bar{\sigma}}_1} \nonumber \\
= \int\frac{d\varepsilon}{2\pi}\frac{\sum_{\lambda}\Gamma^{\lambda}_{\bar{\sigma}} f_{\lambda}(\varepsilon)[1+\frac{i}{2}
\Gamma_{\bar{\sigma}}\langle\langle D_{\bar{\sigma}}|D^{\dagger}_{\bar{\sigma}}\rangle\rangle^{a}_{\varepsilon}]}
{\omega-\varepsilon-\varepsilon_1+i\tilde{\gamma}^{\bar{\sigma}}_1}
\eeq
\beq
\Sigma^{T}_{2\bar{\sigma}}(\omega)=\sum_{\lambda k}\sum_{\lambda' k'} \frac{V_{\lambda k\bar{\sigma}} V^{*}_{\lambda' k'\bar{\sigma}}
 \langle c^{\dagger}_{\lambda k\bar{\sigma}} c_{\lambda' k' \bar{\sigma}}\rangle}
{\omega+\varepsilon_{\lambda k}-\varepsilon_2 +i\tilde{\gamma}_2} \nonumber \\
=\int \frac{d\varepsilon}{2\pi} \frac{\sum_\lambda \Gamma^{\lambda}_{\bar{\sigma}}  f_{\lambda}(\varepsilon) [1-\frac{i}{2}\Gamma_{\bar{\sigma}}\langle\langle D_{\bar{\sigma}}|D^{\dagger}_{\bar{\sigma}}\rangle\rangle^{r}_{\varepsilon}]}{\omega +\varepsilon - \varepsilon_2+ i\tilde{\gamma}_2}
\eeq
The following self-energies:
\be
\Sigma^{(1)}_{\bar{\sigma}}=\sum_{\lambda k}\frac{|V_{\lambda k\bar{\sigma}}|^{2}}{\omega-\varepsilon_{\lambda k}-\varepsilon_{1}+i\tilde{\gamma}^{\bar{\sigma}}_1},
\ee
and
\be
\Sigma^{(2)}_{\bar{\sigma}}=\sum_{\lambda k}\frac{|V_{\lambda k\bar{\sigma}}|^{2}}{\omega+\varepsilon_{\lambda k}-\varepsilon_{2}+i\tilde{\gamma}_2},
\ee
do not depend on the GFs and take on the limiting values $\Sigma_{0\sigma}$ if the inverse lifetimes $\tilde{\gamma}^{\bar{\sigma}}_1$ 
and $\tilde{\gamma}_2$ are positive infinitesimals $0^+$.

We have already seen the relation $\bar{b}_{2\bar{\sigma}}=(1-x)\tilde{b}_{2\bar{\sigma}}$. Even so it is possible
to  find the latter from the former, it is much more convenient to calculate $\tilde{b}_{2\bar{\sigma}}$ directly.
The result reads 
\beq 
\tilde{b}_{2\bar{\sigma}}=(1-x)\int \frac{d\varepsilon}{2\pi} \frac{\sum_{\lambda}\Gamma^{\lambda}_{\bar{\sigma}} f_{\lambda} (\varepsilon)\langle\langle D_{\bar{\sigma}} |D^{\dagger}_{\bar{\sigma}}\rangle\rangle ^{r}_{\varepsilon}}{\omega+\varepsilon -\varepsilon_2 +i\tilde{\gamma}_2} \nonumber \\
+x(2-x)\int \frac{d\varepsilon}{2\pi} \frac{\sum_{\lambda}\Gamma^{\lambda}_{\bar{\sigma}} f_{\lambda} (\varepsilon)\langle\langle n_{\sigma} d_{\bar{\sigma}} |D^{\dagger}_{\bar{\sigma}}\rangle\rangle ^{r}_{\varepsilon}}{\omega+\varepsilon -\varepsilon_2 +i\tilde{\gamma}_2},
\label{tb2bar}
\eeq
and shows that its calculation requires the  
GF $\langle\langle n_{\sigma} d_{\bar{\sigma}} |D^{\dagger}_{\bar{\sigma}}\rangle\rangle ^{r}_{\omega}$. 
It can be easily obtained from (\ref{sol-gf-tn0-1}), and is given by (\ref{sol-gf-tn0-2}).

Other self-energies are expressed in terms of the spectral GF:
\beq
b_{1\bar{\sigma}}(\omega)=\sum_{\lambda k}\frac{V^{*}_{\lambda k\bar{\sigma}}\langle d^{\dagger}_{\bar{\sigma}} c_{\lambda k \bar{\sigma}}
\rangle}{\omega-\varepsilon_1 + i\tilde{\gamma}^{\bar{\sigma}}_1} \nonumber \\ 
= \int \frac{d\varepsilon}{2\pi} \frac{\sum_{\lambda}\Gamma^{\lambda}_{\bar{\sigma}} f_{\lambda}(\varepsilon) \langle\langle D_{\bar{\sigma}}
|d^{\dagger}_{\bar{\sigma}}\rangle\rangle ^{a}_{\varepsilon}}{\omega-\varepsilon -\varepsilon_1 + i\tilde{\gamma}^{\bar{\sigma}}_1}.
\label{b1bs-vs-sgf}
\eeq
In a similar manner one finds: 
\beq
N_{1\bar{\sigma}}(\omega)=\sum_{\lambda k}\frac{V^{*}_{\lambda k\bar{\sigma}}\langle d^{\dagger}_{\bar{\sigma}} n_{\sigma} c_{\lambda k \bar{\sigma}}
\rangle}{\omega-\varepsilon_1 + i\tilde{\gamma}^{\bar{\sigma}}_1} \nonumber \\ 
= \int \frac{d\varepsilon}{2\pi} \frac{\sum_{\lambda}\Gamma^{\lambda}_{\bar{\sigma}} f_{\lambda}(\varepsilon) \langle\langle D_{\bar{\sigma}}
|d^{\dagger}_{\bar{\sigma}}n_{\sigma}\rangle\rangle ^{a}_{\varepsilon}}{\omega-\varepsilon -\varepsilon_1 + i\tilde{\gamma}^{\bar{\sigma}}_1}.
\eeq
The definition $D_\sigma=d_\sigma-xn_{\bar{\sigma}}d_\sigma$ can be used to express $b_{1\bar{\sigma}}$ 
in terms of the spectral GF, 
$\langle\langle d_{\bar{\sigma}}|d^{\dagger}_{\bar{\sigma}}\rangle\rangle ^{a}_{\omega}$, and the related one,
 $\langle\langle n_{\sigma}d_{\bar{\sigma}}|d^{\dagger}_{\bar{\sigma}}\rangle\rangle ^{a}_{\omega}$.

\section{Spectral GF in other decoupling schemes}
\label{app:other-dec}
We have seen that the transport GF is symmetric with respect to $x=1$, while
the spectral one lacks this property. Is this due to the simple decoupling 
procedure (decoupling I) we have applied? As discussed in detail in the Suppl. Material \cite{sm},
there are a number of possibilities to decouple those higher-order GFs which contain products of two lead 
operators. In the paper \cite{vanroermund2010} the GF of the standard Hubbard model has been calculated up to
the order $|V_{\lambda k \sigma}|^4$. Such an approach requires calculations which avoid the decoupling of the GFs with
two lead operators. Similar calculations for the present model are prohibitively difficult, 
thus we stick to the order $|V_{\lambda k \sigma}|^2$. However, even up to this order there is room 
for improvement in relation to the decoupling scheme. The decoupling I  
consists of the most natural projections of the higher-order GFs onto the lower order ones, as explained 
in Eqs.~(\ref{proj-start})--(\ref{proj-fin}) for transport and analogous projections for spectral Green function.  
  
The decoupling II ({\it cf.} Eqs.~(30)--(33)
in the Suppl. Material \cite{sm}) takes into account 
the spin and $x$-dependent shifts of the on-dot energies (Eqs.~(40) and (41) in the Suppl. Material \cite{sm}).
They can be  expressed in terms of two lowest order GFs, see Eq.~(52) in Ref.~\cite{sm}.
This decoupling formally modifies
all self-energies but does not preclude an easy solution for the spectral GF. It still neglects 
the function $\langle\langle n_{\sigma}c_{\lambda k \sigma}|d^{\dagger}_{\sigma}\rangle\rangle^{r}_{\omega}$,
which has not appeared hitherto. This function formally is of the same order as 
$\langle\langle n_{\bar{\sigma}}c_{\lambda k \sigma}|d^{\dagger}_{\sigma}\rangle\rangle^{r}_{\omega}$, and
the hope is that taking it into account restores the symmetry. The inclusion of this GF introduces a 
fourth parameter which we denote $S^{\mathrm{sp}}_\sigma$. With the novel GF, one gets a $4\times 4$ matrix equation for 
the four parameters $S^{\mathrm{sp}}_{n,d,c,\sigma}$. However, it turns out that the inclusion of this 
GF only slightly changes the results, by introducing some asymmetries in both functions even for particle-hole
symmetric systems. In summary, none of the seemingly more involved decouplings, presented 
in the Suppl. Material \cite{sm}, leads to an improvement of the results with respect to the $x$ symmetry.

\section{Matrix formulation}
\label{app:matrix}
As discussed in the Suppl. Material \cite{sm}, one may have another look at the spectral and transport
GFs, stemming from the definition of $D_{\sigma}=d^{\dagger}_{\sigma}(1-xn_{\bar{\sigma}})$. 
This allows to write
\beq
\langle\langle D_{\sigma}|D^{\dagger}_{\sigma}\rangle\rangle_{\omega}&=& 
\langle\langle d_{\sigma}|d^{\dagger}_{\sigma}\rangle\rangle_{\omega} \nonumber \\
&-&x\langle\langle n_{\bar{\sigma}}d_{\sigma}|d^{\dagger}_{\sigma}\rangle\rangle_{\omega} -
x\langle\langle d_{\sigma}|n_{\bar{\sigma}}d^{\dagger}_{\sigma}\rangle\rangle_{\omega} \nonumber \\
&+&x^2\langle\langle n_{\bar{\sigma}}d_{\sigma}|n_{\bar{\sigma}}d^{\dagger}_{\sigma}\rangle\rangle_{\omega},
\eeq
and shows that to get both GFs one has to calculate a matrix GF formally consisting of $d_\sigma$ and
$d_\sigma n_{\bar{\sigma}}$ operators only. 
The matrix GF reads ${\cal G}_\sigma=\langle\langle \phi_{\sigma}|\phi^{\dagger}_{\sigma}\rangle\rangle_{\omega}$ 
where $\phi=\{d_\sigma, n_{\bar{\sigma}}d_\sigma\}^{\cal T}$, and ${\cal T}$ denotes the matrix  transpose operation.
The knowledge of ${\cal G}_\sigma$ is enough to get both spectral and transport GF. This formulation 
seems to be more symmetric compared to that used previously, and thus could, 
in principle at least, lead to the required symmetry not only of the transport but also the spectral GF. 

The details of the calculations and the decouplings are presented in the Suppl. Material \cite{sm}. 
However, we have to stress again that the higher-order GFs appearing in 
all the entries of ${\cal G}_\sigma(\omega)$ have to be approximated.
It turns out that the decoupling I applied to all four components of the matrix GF 
leads to results which are most symmetric and closest to 
the direct calculations presented in App.~\ref{app:trGF}. This is 
illustrated in panel (a) of Fig.~(\ref{fig:rys8}) of the main text for $x=0.5$, and $\varepsilon_d=-5$, $U=8$, $T=0.03$. 
The differences between each set of curves obtained 
by the direct formulae (marked with I), and those obtained with the help of the matrix  
formulation (marked with ``m''), are small, hardly visible for $x=0.5$; but they do depend on $x$ 
and are largest for $x=2$.

Panel (b) of of Fig.~(\ref{fig:rys8}) shows the  spectral and transport densities of states 
for two $x$ values, namely $x=0$ and $x=2$,
calculated by the matrix method. One notices that in the matrix formulations the transport 
density of states for $x=2$ differs from that for $x=0$ also. This is related to the fact that 
in this method not only the charge density $\langle n_\sigma\rangle$ but the full spectral 
GF enters the formulae for the transport GF introducing some asymmetry.    
The comparison of the curves for $N(E)$ for $x=0$ and $x=2$ best illustrate the lack of the required 
symmetry in the spectral function.  One of the main  
problems here is the complete suppression of the Kondo resonance in $g^{r}(\omega)$ for $x=2$. 
While the transport GF for $x=2$ is only quantitatively different from that calculated for $x=0$,
the spectral functions at those two $x$ values are completely different with a minimum (for $x=2$) 
at the chemical potential where the Kondo resonance should appear. 
The source of the asymmetry is discussed in Sec.~\ref{sec:asymmetry}.
The results shown in Fig.~(\ref{fig:rys8}b)
have been obtained from the matrix formulation, using decouplings of all four GFs analogous 
to decoupling I. The other two decouplings  
produce qualitatively similar results. In summary, none of the studied decoupling schemes 
leads to the appearance of the Kondo resonance in the spectral function for $x=2$.

\end{document}